\documentclass[aps,prb,reprint]{revtex4-1}
\usepackage{bm}
\usepackage{amsmath}
\usepackage{here}
\usepackage[dvipdfmx]{graphicx,color}

\def\Vec#1{\mbox{\boldmath $#1$}}

\begin{document}

\title{
  Nitrogen as the best interstitial dopant among $X$=B, C, N, O and F \\
  for strong permanent magnet NdFe$_{11}$Ti$X$: First-principles study
}

\author{Yosuke Harashima,$^{1,3}$ 
  Kiyoyuki Terakura,$^{1,4}$ 
  Hiori Kino,$^{2,3}$ 
  Shoji Ishibashi,$^{1}$ and 
  Takashi Miyake$^{1,3}$} 

\affiliation{
  $^1$Nanomaterials Research Institute, AIST, 
  Tsukuba, Ibaraki 305-8568, Japan\\ 
  $^2$MANA, National Institute for Materials Science, 
  Tsukuba, Ibaraki 305-0044, Japan\\
  $^3$ESICMM, National Institute for Materials Science, 
  Tsukuba, Ibaraki 305-0047, Japan\\
  $^4$National Institute for Materials Science,
  Tsukuba, Ibaraki 305-0047, Japan
}

\date{\today}

\begin{abstract}
  We study magnetic properties of NdFe$_{11}$Ti$X$, 
  where $X$=B, C, N, O, and F, 
  by using first-principles calculations based on 
  density functional theory. 
  Its parent compound NdFe$_{11}$Ti has the ThMn$_{12}$ structure, 
  which has the symmetry of space group $I4/mmm$, No. 139. 
  The magnetization increases 
  by doping B, C, N, O, and F at the $2b$ site of the ThMn$_{12}$ structure. 
  The amount of the increase is larger for $X$=N, O, F than for $X$=B, C. 
  On the other hand, 
  the crystal field parameter $\langle r^{2} \rangle A_{2}^{0}$, 
  which controls the axial magnetic anisotropy of the Nd $4f$ magnetic moment,
  depends differently on the dopant. 
  With increase of the atomic number from $X$=B, 
  $\langle r^{2} \rangle A_{2}^{0}$ increases, 
  takes a maximum value for $X$=N, and then turns to decrease. 
  This suggests that in NdFe$_{11}$Ti$X$, 
  nitrogen is the most appropriate dopant among B, C, N, O, and F 
  for permanent magnets in terms of magnetization and anisotropy. 
  The above calculated properties are explained based on 
  the detailed analysis of the electronic structures of NdFe$_{11}$Ti$X$.
\end{abstract}

\maketitle

\section{Introduction}
\begin{figure}[ht]
  \begin{center}
    \includegraphics[width=0.8\hsize]{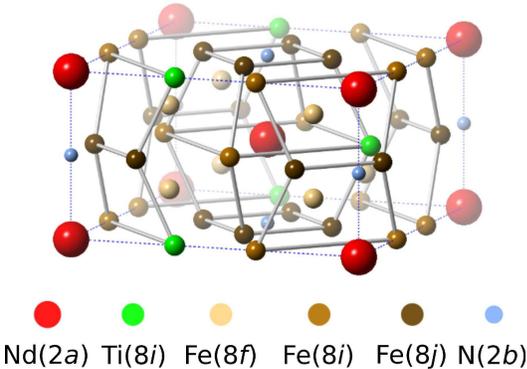}
  \end{center}
  \caption{(Color online) Crystal structure of NdFe$_{11}$TiN. 
    Ti occupies one of the 8$i$ sites.}
  \label{fig:crystalstructure}
\end{figure}

Large remanent magnetization and high coercivity are the two fundamental requirements 
for high performance permanent magnets. 
The remanent magnetization is determined 
by spontaneous spin and orbital moments of a material.
The coercivity is strongly correlated to 
the magnetocrystalline anisotropy energy. 
Extensive studies have been devoted to the search for
higher-performance permanent magnets. 
The strongest permanent magnet to date is based on Nd$_{2}$Fe$_{14}$B, 
which has magnetization of 1.85 T at 4.2 K, 
magnetocrystalline anisotropy field of 67 kOe at room temperature, and 
Curie temperature of 586 K.\cite{HiMaYaFuSaYa1986} 
NdFe$_{11}$TiN (Fig.~\ref{fig:crystalstructure}) is also 
a strong magnet compound, although its maximum energy product is 
smaller than Nd$_{2}$Fe$_{14}$B. 
The interstitial-nitrogenated NdFe$_{11}$Ti has 
magnetization of 1.476 T at 1.5 K, 
magnetocrystalline anisotropy field of 80 kOe at room temperature, and 
Curie temperature of 740 K.\cite{YaZhKoPaGe1991,YaZhGePaKoLiYaZhDiYe1991} 

NdFe$_{11}$Ti has the ThMn$_{12}$ structure (space group $I4/mmm$, No. 139). 
Ideally all of the Mn sites are occupied by Fe. 
However, NdFe$_{12}$ is thermodynamically unstable, 
and some of Fe have to be substituted by another element 
to stabilize the bulk phase (NdFe$_{12-x}M_{x}$). 
For example, Ti, V, Cr, Mn, Mo, W, Al, and Si are known to 
serve as such substitutional elements $M$.\cite{MoBu1988,Bu1991a} 
However, as these substitutional elements reduce 
the magnetization\cite{VeBoZhBu1988}, 
the concentration of the substitutional atoms should be as small as possible. 
The concentration range of the substitution, $x$, depends on 
the element. 
In this regard, 
titanium is a favorable element because it stabilizes the material with 
small $x$ ($\approx 1$). 

To optimize the performance of a strong magnet, 
we can utilize even another degree of freedom, i.e., the interstitial dopant. 
In fact, magnetic properties of NdFe$_{12-x}M_{x}$ are significantly 
controlled by interstitial doping of a typical element.
It is experimentally observed that 
the interstitial nitrogenation enhances the magnetic moment 
from 21.273 (NdFe$_{11}$Ti) to 23.218 $\mu_{B}$/f.u. 
(NdFe$_{11}$TiN$_{0.5}$).\cite{YaZhKoPaGe1991} 
The Curie temperature ($T_{\mathrm{C}}$) also rises from 570 K in NdFe$_{11}$Ti 
to 740 K in its nitride.\cite{YaZhKoPaGe1991,YaZhGePaKoLiYaZhDiYe1991} 
Furthermore, strong uniaxial magnetocrystalline anisotropy is induced 
by the nitrogenation.
\cite{YaZhKoPaGe1991,YaZhGePaKoLiYaZhDiYe1991,AkFuYaTa1994}
Another possible interstitial dopant is carbon.
\cite{HuCo1992,LiZhMo1993,YaOlEcWoMu2000} 
Carbon doping at an interstitial site leads to 
increase of $T_{\mathrm{C}}$ and uniaxial anisotropy 
in NdFe$_{12-x}M_{x}$ compounds. 
However, nitrogenation is more preferable than carbonation 
in terms of magnetization and Curie temperature. 
In order to design better strong magnets, 
the cause of the difference between N and C has to be clarified and the possibility of 
better dopants has to be searched for. 

In previous papers,~\cite{MiTeHaKiIs2014,HaTeKiIsMi2015a}
we have theoretically studied the effects of Ti substitution and nitrogenation 
by comparing NdFe$_{12}$, NdFe$_{11}$Ti and NdFe$_{11}$TiN using experimental structures. 
We found that Ti substitution in NdFe$_{12}$ reduces the magnetic moment 
more than subtracting the local magnetic moment at the substituted Fe site, 
and enhances the uniaxial anisotropy slightly. 
It was also shown that interstitial nitrogenation increases 
the magnetic moment by 2.75 $\mu_{B}$/f.u. in NdFe$_{11}$Ti. 
The magnetic anisotropy of NdFe$_{11}$Ti and its nitride 
was studied by using the crystal field parameter 
$\langle r^{2} \rangle A_{2}^{0}$. 
We found that the interstitial nitrogenation substantially increases 
$\langle r^{2} \rangle A_{2}^{0}$. 
This implies that 
the uniaxial anisotropy is enhanced by the nitrogenation in NdFe$_{11}$Ti. 
Therefore, the interstitial nitrogenation works preferably in terms of 
magnetization and magnetocrystalline anisotropy. 

Stimulated by our work, the experimental group tried to synthesize 
a new strong magnet compound NdFe$_{12}$N and succeeded 
in forming a film recently.\cite{HiTaHiHo2014}
It has been shown that the film sample has better intrinsic magnetic properties 
than Nd$_2$Fe$_{14}$B, 
although a bulk sample is still difficult to synthesize. 
This experimental work, in turn, motivated us to revisit 
NdFe$_{11}$TiN and related compounds 
in order to search for stronger permanent magnet compounds. 

In the present paper, 
we perform detailed electronic structure calculations for NdFe$_{11}$Ti$X$ 
with $X$ the interstitial impurity of typical elements B, C, N, O, and F. 
We analyze in detail the electronic structures, paying particular attention to 
the $X$-$2p$ states.  
Then, based on the analysis we explain the variations in structural properties, 
magnetization, and magnetocrystalline anisotropy caused by 
the interstitial dopant $X$. 

\section{Calculation methods}
The calculations are carried out by using the first principles code 
QMAS (Quantum MAterials Simulator)\cite{Qm2014} 
which is based on density functional theory\cite{HoKo1964,KoSh1965} and 
the projector augmented-wave (PAW) method.\cite{Bl1994,KrJo1999}
For the exchange-correlation energy functional, 
the generalized gradient approximation (GGA) is used.\cite{PeBuEr1996}
The $8 \times 8 \times 8$ k points are sampled and 
the cutoff energy for the plane wave basis is set to 40.0 Ry. 
The lattice constants and inner coordinates are optimized. 
NdFe$_{11}$Ti$X$ has $4f$ orbitals 
that are strongly localized at the Nd sites. 
In the present study, the $f$-states of Nd are treated as open-core states,  
in which the hybridization with other orbitals is neglected completely 
and the atomic physics is applied to the Nd-$4f$ states.
Following Hund's first rule, we assume that 
Nd has three $f$-electrons with the full spin polarization. 
The self-consistent calculation makes the $4f$ spins antiparallel to 
the Fe spins. 

The orbital magnetic moment is not treated in 
our self-consistent calculations, 
but that of the Nd-$4f$ electrons is included 
in the total magnetic moment as follows. 
We assume that the Nd-$4f$ electrons yield the local magnetic moment of $g_{J}J$, 
where $g_{J}$ is the Lande g-factor, and $J=9/2$ is the total angular momentum 
for the Nd-$4f$ electrons. 
(For Nd-$4f$ electrons, $J=|L-S|$, 
the orbital angular momentum $L=6$, and 
the spin angular momentum $S=3/2$ are given by Hund's rules.)
The total magnetic moment of the whole material is estimated by 
adding $g_{J}J$=3.273 $\mu_{B}$ to the spin magnetic moment.
This spin magnetic moment is the contribution from 
the valence electrons other than the Nd-$4f$ electrons. 
Fe has only small spin-orbit interaction compared with Nd; 
thus the spin-orbit interaction on Fe is neglected.

The magnetocrystalline anisotropy energy is expressed in terms of 
a first-order coefficient $K_{1}$ as
\begin{equation}
  \label{mae}
  E(\theta) \approx K_{1} \sin^{2} \theta \;,
\end{equation}
where $\theta$ is the rotation angle of the magnetization measured 
from the $c$ axis.
We focus on 
the contribution of the rare-earth $4f$ electrons. 
It can be evaluated by the interaction energy between the $4f$ electrons and 
the surrounding crystalline electric field. 
Because of Hund's rules, 
the electron distribution of the Nd-$4f$ states 
deviates significantly from the spherical symmetry. 
The shape of the electron density distribution is uniquely related to
$\Vec{J}$ which is parallel to the $4f$ magnetic moment. 
The non-spherical electron density distribution couples with 
the crystal-electric field leading to the magnetic anisotropy. 
Following the crystal-field theory, 
$K_{1}$ can be expressed 
by using the crystal field parameter $A_{2}^{0}$ as 
\begin{equation}
  \label{eq:mae-cef}
  K_{1} = -3J(J-\frac{1}{2}) \alpha_{J} \langle r^{2} \rangle A_{2}^{0} n_{R} \;.
\end{equation}
Here, $\alpha_{J}$ is the Stevens factor
of the rare-earth atoms, which is determined for each rare-earth element. 
For Nd$^{3+}$, $\alpha_{J}=-7/1089$.
$n_{R}$ is the Nd concentration.
We evaluate $\langle r^{2} \rangle A_{2}^{0}$ by the equation
\begin{equation}
  \label{eq:cef_veff}
  \langle r^{l} \rangle A_{l}^{m} =
  F_{l}^{m} \int_{0}^{r_{c}} W_{l}^{m} \left( r \right) 
  \phi^{2} \left( r \right) dr \;.
\end{equation}
$F_{l}^{m}$ is a prefactor of the real spherical harmonics $Z_{l}^{m}$ 
and its explicit expressions can be found in, e.g., 
Ref.~\onlinecite{RiOpEsJo1992}. 
$W_{l}^{m}$ is the effective potential at the Nd site expanded by $Z_{l}^{m}$. 
$\phi$ is the radial function of the Nd-$4f$ orbital, 
which is obtained in GGA with the self interaction correction. 
In Eq.~(\ref{eq:mae-cef}) 
it is implicitly expected that $\phi$ is well localized, 
and we neglect the contribution from a tail of $\phi$ 
in Eq.~(\ref{eq:cef_veff}) 
by introducing a cutoff radius $r_{c}$.
The cutoff radius is determined so that the volume of the sphere is equal to 
that of the Bader region.~\cite{Ba1990,HeArJo2006}
(See Fig. S1 of the Supplemental Material~\cite{supplementalmaterial} 
for the actual value of the atomic sphere for each $X$. 
The atomic radius for NdFe$_{11}$Ti$X$ is also used for NdFe$_{11}$Ti$E_{X}$, 
which denotes NdFe$_{11}$Ti given by removing $X$ from NdFe$_{11}$Ti$X$ 
with other atoms fixed at their positions in the structure optimized.)

\begin{table}
  \caption{Optimized lattice constants of NdFe$_{11}$Ti$X$ in units of \AA. 
    Due to Ti substitution, 
    the system is no longer tetragonal but orthorhombic  ($a \neq b$).}
  \vspace{5pt}
  \begin{tabular}{r||@{\hspace{10pt}}c@{\hspace{10pt}}|
      @{\hspace{10pt}}c@{\hspace{10pt}}|@{\hspace{10pt}}c@{\hspace{10pt}}|c}
    & $a$ & $b$ & $c$ & volume \\
    \hline
    \hline
    $X$=B & 8.521 & 8.593 & 4.943 & 180.9 \\
    \hline
    C     & 8.495 & 8.572 & 4.914 & 178.9 \\
    \hline
    N     & 8.537 & 8.618 & 4.880 & 179.5 \\
    \hline
    O     & 8.658 & 8.704 & 4.786 & 180.3 \\
    \hline
    F     & 8.812 & 8.830 & 4.745 & 184.6 \\
    \hline
    empty & 8.553 & 8.568 & 4.701 & 172.3 \\
  \end{tabular}
  \label{table:latticeconstant}
\end{table}
\begin{table*}
  \caption{Optimized inner coordinates. The Nd atom is fixed at the origin.
    Even though the symmetry of the ThMn$_{12}$ structure 
    is broken by the Ti substitution, 
    we use the notation Fe($8f$), Fe($8i$), and Fe($8j$) 
    in NdFe$_{11}$Ti$X$.} 
  \vspace{5pt}
  \begin{tabular}{r||c|c|c|c}
    & Fe(8$f$) & Fe(8$i$) & Fe(8$j$) & $X$ \\ 
    \hline
    \hline
    $X$=B & ( 0.257, 0.252, 0.250 ) & (  0.374,  0.000, 0.000 ) (Ti) & (  0.274,  0.500, 0.000 ) & ( 0.006, 0.000, 0.500 ) \\
          & ( 0.257, 0.748, 0.750 ) & ( -0.351,  0.000, 0.000 )      & ( -0.262,  0.500, 0.000 ) & \\
          & ( 0.757, 0.248, 0.750 ) & (  0.007,  0.359, 0.000 )      & (  0.510,  0.272, 0.000 ) & \\
          & ( 0.757, 0.752, 0.250 ) & (  0.007, -0.359, 0.000 )      & (  0.510, -0.272, 0.000 ) & \\
    \hline
    C     & ( 0.256, 0.252, 0.250 ) & (  0.374,  0.000, 0.000 ) (Ti) & (  0.281,  0.500, 0.000 ) & ( 0.005, 0.000, 0.500 ) \\
          & ( 0.256, 0.748, 0.750 ) & ( -0.351,  0.000, 0.000 )      & ( -0.269,  0.500, 0.000 ) & \\
          & ( 0.756, 0.248, 0.750 ) & (  0.007,  0.360, 0.000 )      & (  0.509,  0.279, 0.000 ) & \\
          & ( 0.756, 0.752, 0.250 ) & (  0.007, -0.360, 0.000 )      & (  0.509, -0.279, 0.000 ) & \\
    \hline
    N     & ( 0.256, 0.252, 0.250 ) & (  0.373,  0.000, 0.000 ) (Ti) & (  0.279,  0.500, 0.000 ) & ( 0.005, 0.000, 0.500 ) \\
          & ( 0.256, 0.748, 0.750 ) & ( -0.352,  0.000, 0.000 )      & ( -0.269,  0.500, 0.000 ) & \\
          & ( 0.756, 0.248, 0.750 ) & (  0.006,  0.360, 0.000 )      & (  0.510,  0.277, 0.000 ) & \\
          & ( 0.756, 0.752, 0.250 ) & (  0.006, -0.360, 0.000 )      & (  0.510, -0.277, 0.000 ) & \\
    \hline
    O     & ( 0.256, 0.251, 0.250 ) & (  0.373,  0.000, 0.000 ) (Ti) & (  0.271,  0.500, 0.000 ) & ( 0.003, 0.000, 0.500 ) \\
          & ( 0.256, 0.749, 0.750 ) & ( -0.351,  0.000, 0.000 )      & ( -0.263,  0.500, 0.000 ) & \\
          & ( 0.756, 0.249, 0.750 ) & (  0.006,  0.360, 0.000 )      & (  0.509,  0.270, 0.000 ) & \\
          & ( 0.756, 0.751, 0.250 ) & (  0.006, -0.360, 0.000 )      & (  0.509, -0.270, 0.000 ) & \\
    \hline
    F     & ( 0.257, 0.251, 0.250 ) & (  0.373,  0.000, 0.000 ) (Ti) & (  0.252,  0.500, 0.000 ) & ( 0.005, 0.000, 0.500 ) \\
          & ( 0.257, 0.749, 0.750 ) & ( -0.349,  0.000, 0.000 )      & ( -0.246,  0.500, 0.000 ) & \\
          & ( 0.757, 0.249, 0.750 ) & (  0.006,  0.358, 0.000 )      & (  0.510,  0.253, 0.000 ) & \\
          & ( 0.757, 0.751, 0.250 ) & (  0.006, -0.358, 0.000 )      & (  0.510, -0.253, 0.000 ) & \\
    \hline
    empty & ( 0.254, 0.251, 0.251 ) & (  0.374,  0.000, 0.000 ) (Ti) & (  0.269,  0.500, 0.000 ) & \\
          & ( 0.254, 0.749, 0.749 ) & ( -0.350,  0.000, 0.000 )      & ( -0.264,  0.500, 0.000 ) & \\
          & ( 0.754, 0.249, 0.751 ) & (  0.005,  0.358, 0.000 )      & (  0.506,  0.274, 0.000 ) & \\
          & ( 0.754, 0.751, 0.249 ) & (  0.005, -0.358, 0.000 )      & (  0.506, -0.274, 0.000 ) & 
  \end{tabular}
  \label{table:coordinate}
\end{table*}
\begin{table}
  \caption{Magnetic moment $m$ [$\mu_{\mathrm{B}}$/f.u.], 
    magnetization $\mu_{0}M$ [T], 
    crystal field parameter $\langle r^{2} \rangle A_{2}^{0}$ [K], 
    magnetocrystalline anisotropy energy coefficient $K_{1}$ [MJ/m$^{3}$], 
    and anisotropy field $\mu_{0}H_{a}$ [T] 
    of NdFe$_{11}$Ti$X$ are shown 
    for $X=$B, C, N, O, F, and NdFe$_{11}$Ti.}
  \vspace{5pt}
  \begin{tabular}{r||c|c|c|c|c}
    & $m$ & $\mu_{0}M$ & $\langle r^{2} \rangle A_{2}^{0}$ & $K_{1}$ & $\mu_{0}H_{a}$ \\
    & [ $\mu_{\mathrm{B}}$/f.u. ] & \hspace{7pt} [ T ] \hspace{7pt}
    & \hspace{6pt} [ K ] \hspace{6pt} & [ MJ/m$^{3}$ ] 
    & \hspace{6pt} [ T ] \hspace{6pt} \\
    \hline
    \hline
    $X$=B & 25.51 & 1.643 & -27  & -0.70 & -1.1  \\
    \hline
    C     & 25.18 & 1.640 &  96  &  2.6  &  4.0  \\
    \hline
    N     & 26.84 & 1.743 &  397 & 10.6  &  15.3 \\
    \hline
    O     & 27.23 & 1.760 & -7   & -0.2  & -0.3  \\
    \hline
    F     & 27.07 & 1.709 & -502 & -13.0 & -19.2 \\
    \hline
    empty & 24.10 & 1.631 & -29  & -0.80 & -1.2  \\
  \end{tabular}
  \label{table:magproperty}
\end{table}

\section{Results}
\begin{figure}[ht]
  \begin{center}
    \includegraphics[width=\hsize]{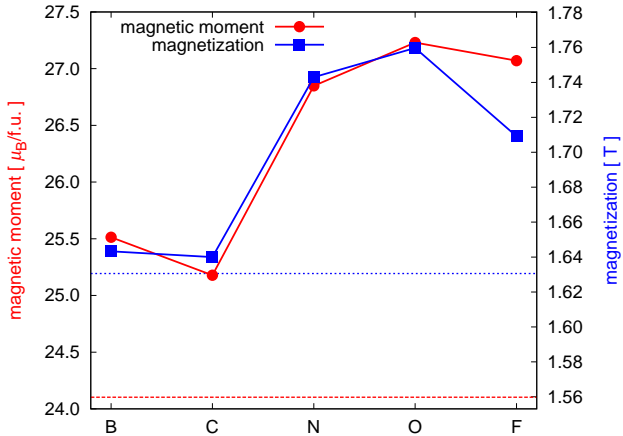}
  \end{center}
  \caption{(Color online) 
    Total magnetic moment and magnetization of NdFe$_{11}$Ti$X$ 
    ($X$=B, C, N, O, F) in units of 
    $\mu_{B}$/f.u. (left axis) and tesla (right axis). 
    Theoretically optimized lattice constants and inner coordinates are used. 
    In all the dopants, the magnetization is larger than 
    that of NdFe$_{11}$Ti, 1.631 T, shown as a blue dotted line 
    (corresponding magnetic moment is 24.10 $\mu_{B}$/f.u. 
    shown as a red broken line). 
  }
  \label{fig:mag_total}
\end{figure}
\begin{figure}[ht]
  \begin{center}
    \includegraphics[width=\hsize]{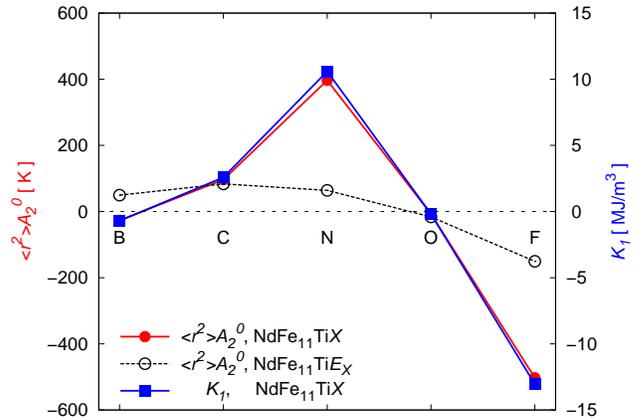}
  \end{center}
  \caption{(Color online) $X$ dependence of 
    $\langle r^{2} \rangle A_{2}^{0}$ and $K_{1}$. 
    The filled red circles and empty black circles correspond to 
    $\langle r^{2} \rangle A_{2}^{0}$ of 
    NdFe$_{11}$Ti$X$ and NdFe$_{11}$Ti$E_{X}$, respectively. 
    The latter is 
    the system without $X$ keeping the other atoms fixed at 
    the positions of the former.
    The lattice constant and inner coordinate are optimized for 
    NdFe$_{11}$Ti$X$. 
    (The parameters are shown in 
    Tables~\ref{table:latticeconstant} and \ref{table:coordinate}.)
    The structural effect shown by empty black circles is not dominant 
    in the dopant dependence of $\langle r^{2} \rangle A_{2}^{0}$. 
    The details are discussed in Sec.~\ref{sec:crystalfieldparameter}. 
    In addition, $K_{1}$ is 
    estimated from $\langle r^{2} \rangle A_{2}^{0}$ 
    by using Eq.~(\ref{eq:mae-cef}), 
    and shown as the blue squares. 
    The red circles and blue squares do not scale exactly 
    because of the volume effect through $n_R$ in Eq.~(\ref{eq:mae-cef}).
  }
  \label{fig:cef_total}
\end{figure}
From the total energy of NdFe$_{11}$Ti, 
we found that Ti substitution is more stable at the 8$i$ site than 
the 8$f$ and 8$j$ sites by 0.78 and 0.51 eV/f.u., respectively, 
which agrees with experimental indication.
\cite{YaZhGePaKoLiYaZhDiYe1991,MoIbBu1991} 
(Even though the symmetry of the ThMn$_{12}$ structure is broken by 
the Ti substitution, 
we use the notation of ThMn$_{12}$ for the structure of NdFe$_{11}$Ti$X$.) 
Hereafter, Ti is put at the 8$i$ site and 
$X$ is inserted at the midpoint between Nd atoms 
along the $c$ axis as shown in Fig.~\ref{fig:crystalstructure}.
%
The optimized lattice constants and fractional coordinates for 
NdFe$_{11}$Ti and NdFe$_{11}$Ti$X$ are shown 
in Tables~\ref{table:latticeconstant} and \ref{table:coordinate}, respectively, 
where "empty" denotes NdFe$_{11}$Ti. 
The interstitial $X$ expands the volume by 
5\% for $X$=B, 4\% for C, 4\% for N, 5\% for O, and 7\% for F. 

In the present paper, 
we distinguish between "magnetic moment" $m$ and "magnetization" $\mu_{0}M$. 
"magnetic moment" is expressed in units of the Bohr magneton $\mu_{B}$ 
per atom or per formula unit, 
while "magnetization" is in units of Tesla 
which is estimated from $m$, volume, and the vacuum permeability $\mu_{0}$. 
$K_{1}$ is evaluated from the obtained $\langle r^{2} \rangle A_{2}^{0}$ 
by using Eq.~(\ref{eq:mae-cef}). 
We also estimate the magnetocrystalline anisotropy field $\mu_{0}H_{a}$ 
($\equiv 2\mu_{0}K_{1}/\mu_{0}M$). 
The calculated magnetic properties are summarized 
in Table~\ref{table:magproperty}. 

The total magnetic moments are calculated to be 
24.10 $\mu_{B}$/f.u. for NdFe$_{11}$Ti and 26.84 $\mu_{B}$/f.u. for 
NdFe$_{11}$TiN with the structure optimization for each system. 
They correspond to the magnetization of 1.631 T and 1.743 T, respectively.
In the previous paper, we reported the spin magnetic moment 
for the experimental lattice parameters. \cite{MiTeHaKiIs2014} 
The results were 19.99 and 20.97 $\mu_{B}$/f.u. 
for NdFe$_{11}$Ti and NdFe$_{11}$TiN, respectively. 
Here, the spin magnetic moments of the valence electrons other than 
that of the Nd-$4f$ electrons are 22.99 and 23.97 $\mu_{B}$/f.u. 
The corresponding total magnetic moments can be estimated 
by adding $g_{J}J$=3.273 $\mu_{B}$ to these spin magnetic moment as 
26.26 $\mu_{B}$/f.u. (1.697 T) and 27.24 $\mu_{B}$/f.u. (1.731 T), 
respectively.
The nitrogenation enhances the magnetization for 
both optimized and experimental lattice parameters, 
and the increment is larger in the former case (0.112 T) than 
the latter (0.034 T). 
Experimentally, the magnetic moment is 
21.273 $\mu_{B}$/f.u. in NdFe$_{11}$Ti and 
23.218 $\mu_{B}$/f.u. in NdFe$_{11}$TiN$_{0.5}$,~\cite{YaZhKoPaGe1991} 
which correspond to 1.375 T and 1.476 T. 
The increment of the magnetic moment and magnetization are 
3.89 $\mu_{B}$ and 0.202 T per nitrogen, respectively. 
In Ref.~\onlinecite{AkFuYaTa1994}, on the other hand, 
the magnetizations of NdFe$_{11}$Ti and NdFe$_{11}$TiN$_{1.5}$ are 
reported as 1.70 T and 1.84 T, respectively; thus  
the magnetization difference is 0.093 T per nitrogen. 
These experimental results are in reasonable agreement 
with our calculation for the optimized lattice parameters. 

Figure~\ref{fig:mag_total} shows the magnetic moment and magnetization 
for $X$=B, C, N, O, and F. 
The magnetic moment is increased by all the interstitial dopants. 
The corresponding magnetization is also larger than that 
for NdFe$_{11}$Ti (1.631 T) 
but the amount of increase is partially canceled by the volume expansion. 
The increment of the magnetization for $X$=B and C is small, 
while that for $X$=N, O, and F is significant with  a jump between $X$=C and N. 
Consequently, the interstitial doping of N, O, and F 
works positively for permanent magnets than doping of B and C
in terms of the magnetization. 

We now turn to the discussion on magnetocrystalline anisotropy based on 
the lowest crystal field parameter $A_{2}^{0}$ at the Nd site. 
Since Nd has negative Stevens factor, 
positive (negative) value of $A_{2}^{0}$ implies 
uniaxial (in-plane) anisotropy. 
In the case of $X$=N, 
the interstitial doping increases the value of 
$\langle r^{2} \rangle A_{2}^{0}$ from $-$29 K to 397 K. 
According to Eq.~(\ref{eq:mae-cef}), 
these values correspond to $K_{1}=-0.80$ MJ/m$^{3}$ and 
10.6 MJ/m$^{3}$, respectively. 
Thus, strong uniaxial anisotropy by nitrogenation 
is indicated from the calculation. 
The experimental anisotropy energy constant 
can be estimated from the magnetization and anisotropy field 
as $K_{1}=6.75$ MJ/m$^{3}$ in 
NdFe$_{11}$TiN$_{0.5}$ at 1.5 K~\cite{YaZhKoPaGe1991} 
or larger than 18 MJ/m$^{3}$ in 
NdFe$_{11}$TiN$_{1.5}$ at 4.2 K~\cite{AkFuYaTa1994}. 
If we simply assume a linear relation between dopant concentration and 
the increment of $K_{1}$, 
our theoretical estimate is 
4.9 MJ/m$^3$ for the former with N$_{0.5}$ and 
16.3 MJ/m$^3$ for the latter with N$_{1.5}$. 

The interstitial dopant dependence 
of $\langle r^{2} \rangle A_{2}^{0}$ is shown in Fig.~\ref{fig:cef_total}.
$\langle r^{2} \rangle A_{2}^{0}$ is $-$27 K for $X$=B. 
With increasing atomic number, 
$\langle r^{2} \rangle A_{2}^{0}$ increases up to $X$=N, 
then turns to decrease, and takes a large negative value for $X$=F. 
Thus, the N doping is suggested to induce the strongest uniaxial anisotropy 
among the typical elements, B, C, N, O, and F.

\section{Discussion}
\subsection{Electronic Structures}
\subsubsection{Fe-$3d$ band}
\begin{figure}[ht]
  \begin{center}
    \includegraphics[width=\hsize]{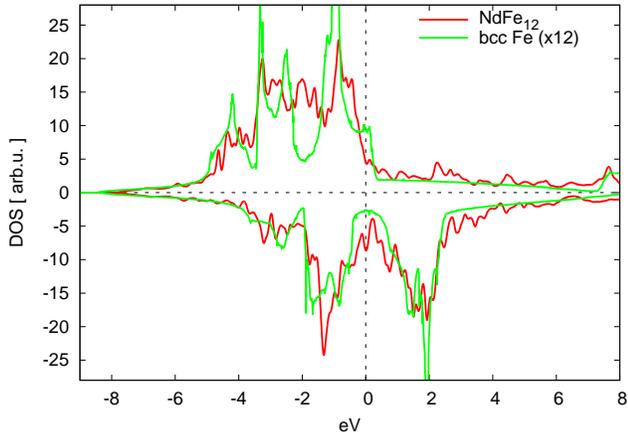}
  \end{center}
  \caption{(Color online) Density of states for NdFe$_{12}$ (red line) and 
    bcc Fe (green line).
    DOS of bcc Fe has a shoulder in the majority-spin band at the Fermi level, 
    and the width of DOS of NdFe$_{12}$ is narrower than that of bcc Fe.
    The origin of the energy is set at the Fermi level.}
  \label{fig:dos_bcc-fe_ndfe12}
\end{figure}
As our analysis of 
the previous paper~\cite{MiTeHaKiIs2014} and 
the present work depends on the fact that the Fermi level lies above 
the majority-spin Fe-$3d$ band, 
we first confirm this fact by studying the density of states (DOS). 
Figure~\ref{fig:dos_bcc-fe_ndfe12} shows DOSs of both of 
NdFe$_{12}$ and bcc Fe. 
Because of the presence of a shoulder of DOS for bcc Fe at the Fermi level 
in the majority-spin state, 
the width of the majority-spin Fe-$3d$ band looks to be slightly narrower 
in NdFe$_{12}$ than that in bcc Fe. 
In order to make the comparison more quantitative, 
we estimated the second moment of the Fe-$3d$ band DOS with 
the tight-binding picture using the neighboring Fe-Fe distances and 
the coordination numbers. 
The inverse fifth power law was assumed for 
the interatomic distance dependence of 
the $d$-$d$ hopping integral predicted 
by the canonical band picture.\cite{An1975} 
We found that the second moment for NdFe$_{12}$ is 0.91 of that for bcc Fe 
being consistent with the difference in the band width mentioned above. 
This implies that the effective Fe-Fe distance is elongated in NdFe$_{12}$ 
compared with that in bcc Fe. 
The narrower $d$ band width of NdFe$_{12}$ 
makes the majority-spin Fe-$3d$ band virtually filled, 
and is expected to give larger magnetic moment. 
The spin moment per Fe is 
2.21 $\mu_{\mathrm{B}}$ in NdFe$_{12}$ and 2.18 $\mu_{\mathrm{B}}$ in bcc Fe. 
This is consistent with the description above. 
However, the ratio of the spin moments $2.18/2.21=0.99$ is close to $1$, 
whereas that of the second moment is 0.91. 
In NdFe$_{12}$ there is also the hybridization between Nd-$5d$ and Fe-$3d$.
As Nd-$5d$ states are located above Fe-$3d$ states, 
this additional hybridization pushes down 
the minority-spin state more than 
the majority-spin state of Fe-$3d$ and 
reduces the magnetic moment in NdFe$_{12}$, 
which explains such small difference in the spin moment 
between NdFe$_{12}$ and bcc Fe. 
The situation is not basically changed in NdFe$_{11}$Ti 
except the appearance of the split-off Ti-$3d$ states.\cite{MiTeHaKiIs2014}

\subsubsection{$X$-$2s$ and $2p$ states}
\begin{figure}[ht]
  \begin{center}
    \includegraphics[width=\hsize]{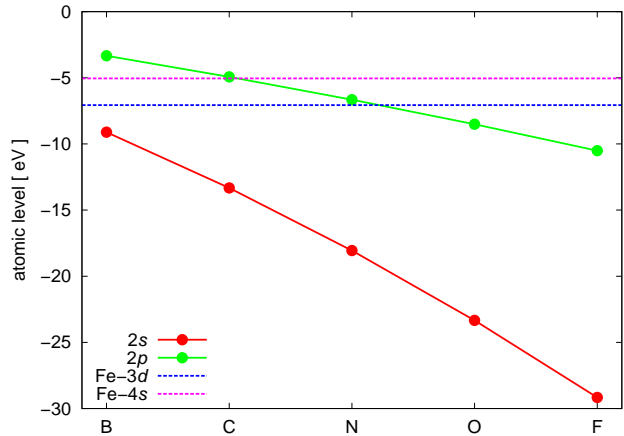}
  \end{center}
  \caption{(Color online) 
  Atomic eigen levels of $X$-$2p$ and -$2s$ states are shown 
  for $X=$B, C, N, O, and F. 
  Fe-$3d$ and -$4s$ levels are also shown.
  The vacuum level is taken as the origin of the vertical axis.}
  \label{fig:atomic_level}
\end{figure}
\begin{figure}[ht]
  \begin{center}
    \includegraphics[width=0.6\hsize]{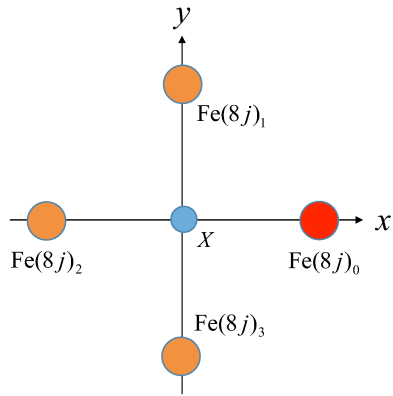}
  \end{center}
  \caption{(Color online)
    The local coordinate defined with respect to the pair between 
    a particular $X$ (blue circle) and a particular Fe(8$j$) (red circle).
  }
  \label{fig:relativeaxis_fe8j}
\end{figure}
\begin{figure}[ht]
  \begin{center}
    \includegraphics[width=\hsize]{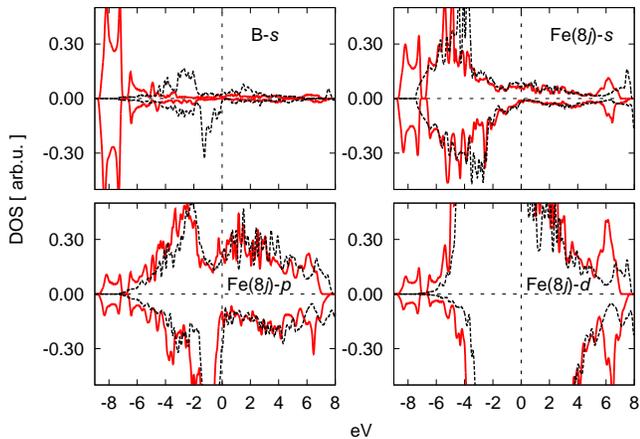}
  \end{center}
  \caption{(Color online) Partial density of states of 
    NdFe$_{11}$TiB (red solid line) and 
    NdFe$_{11}$Ti$E_{X}$ (black broken line).
    The B-$s$, Fe(8$j$)-$s$, -$p$, -$d$ components are shown.
    The origin of the energy is set at the Fermi level.
    The split-off states can be seen around $-$8eV in the B-$s$ component. 
    Note that the weight around $+$6eV in the Fe(8$j$) components is not 
    the antibonding states coupled with B-$s$ states, 
    but the antibonding states coupled with B-$p$ states.}
  \label{fig:dos_splitoff}
\end{figure}
\begin{figure*}[ht]
  \begin{center}
    \includegraphics[width=\hsize]{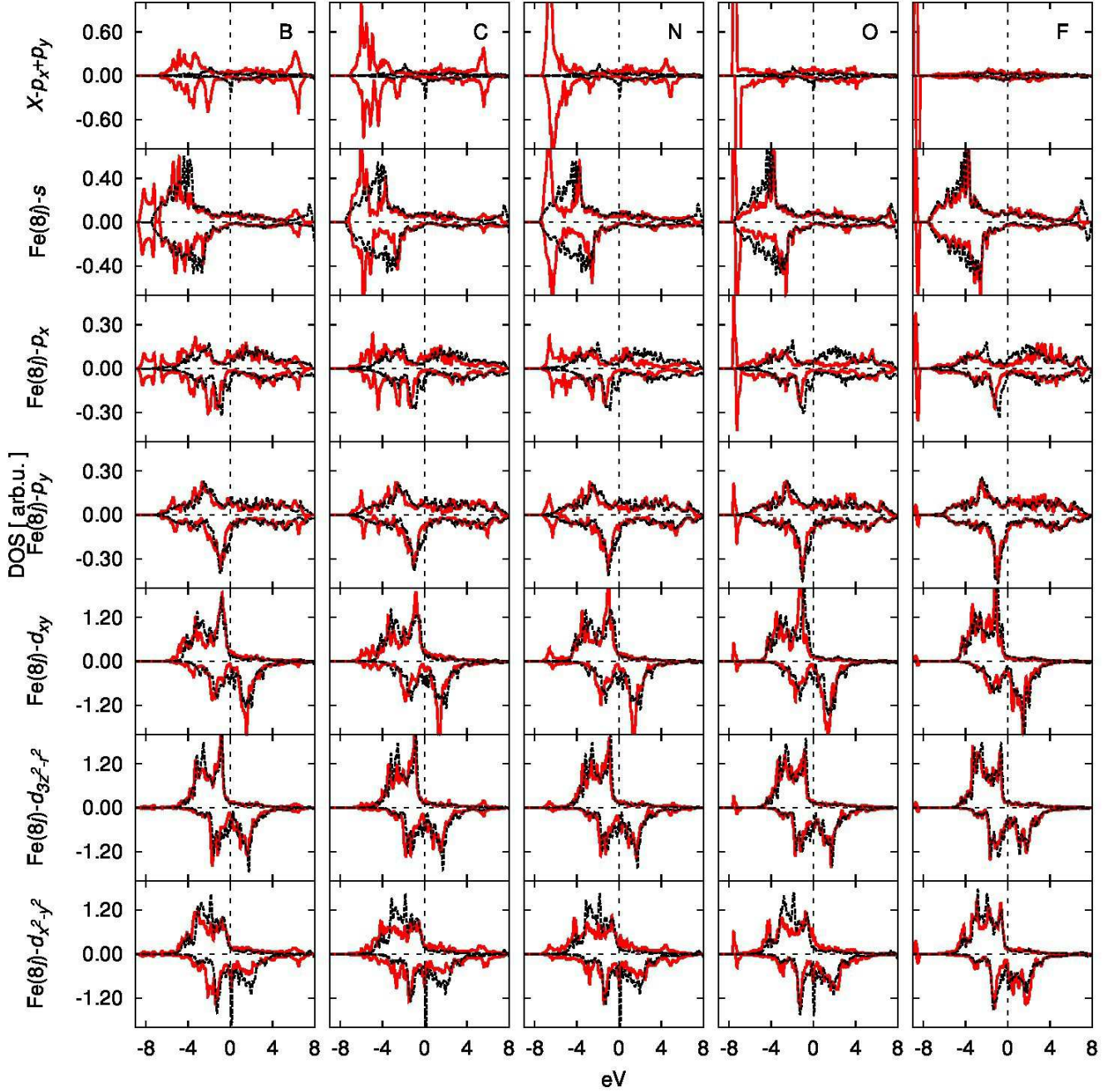}
  \end{center}
  \caption{(Color online) 
    The partial DOS relating to the $\sigma$ bond between 
    $X$-$p$ and Fe(8$j$) states 
    [$X$-$p_{x}$, $p_{y}$, 
    Fe(8$j$)-$s$, $p_{x}$, $d_{3z^{2}-r^{2}}$, $d_{x^{2}-y^{2}}$]. 
    Moreover, Fe(8$j$)-$p_{y}$ and $d_{xy}$ are also shown. 
    The red solid lines are for NdFe$_{11}$Ti$X$ and
    the black broken lines are for NdFe$_{11}$Ti$E_{X}$.
    The origin of the energy is set at the Fermi level.}
  \label{fig:dos_x-sigma}
\end{figure*}
\begin{figure*}[ht]
  \begin{center}
    \includegraphics[width=\hsize]{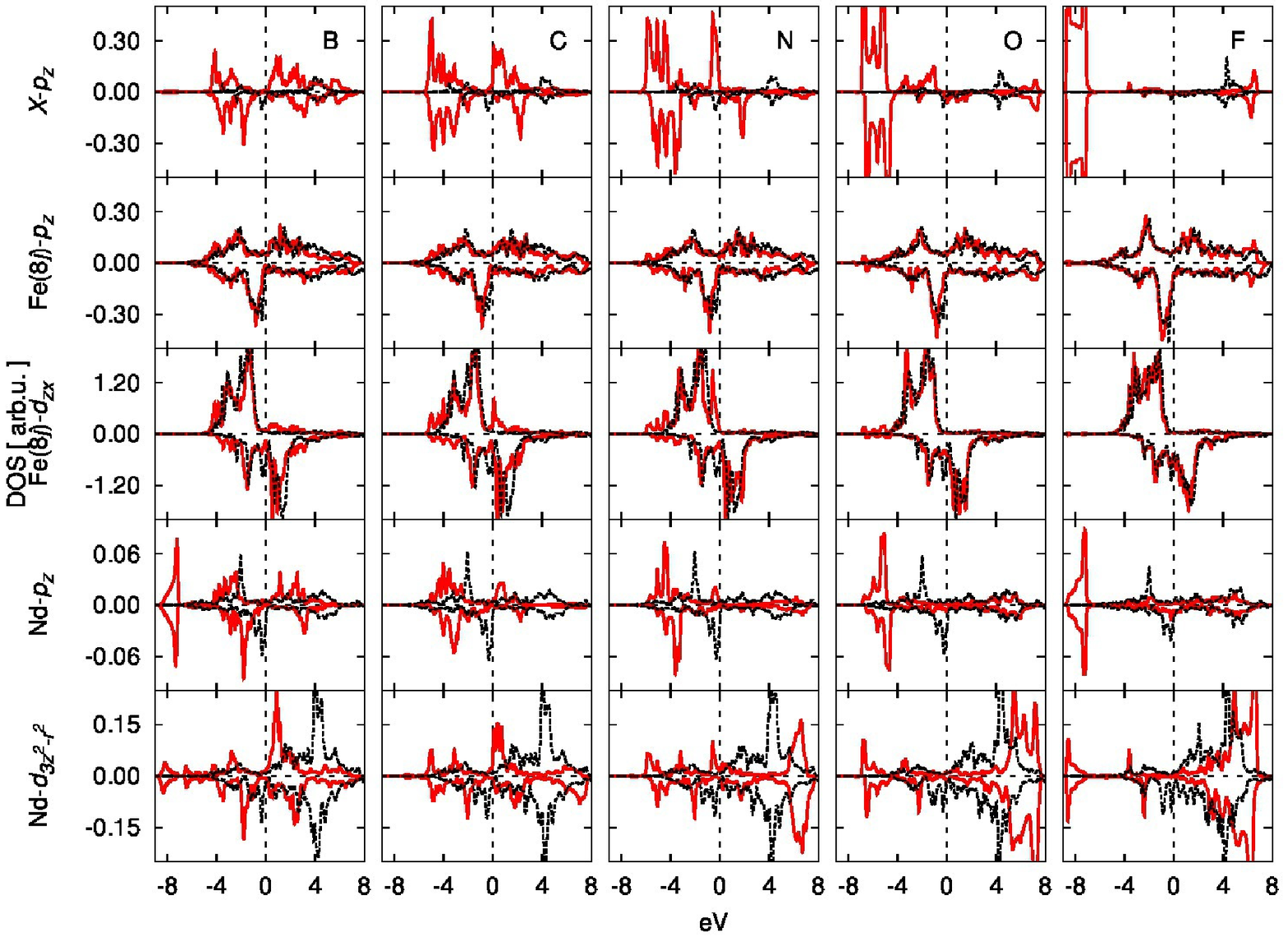}
  \end{center}
  \caption{(Color online) 
    The partial DOS relating to the $\pi$ bond between 
    $X$-$p$ and Fe(8$j$) states 
    [$X$-$p_{z}$, Fe(8$j$)-$p_{z}$, $d_{zx}$].
    Nd-$p_{z}$ and $d_{3z^{2}-r^{2}}$ are also shown.
    The red solid lines are for NdFe$_{11}$Ti$X$ and
    the black broken lines are for NdFe$_{11}$Ti$E_{X}$. 
    The origin of the energy is set at the Fermi level.} 
  \label{fig:dos_x-pi}
\end{figure*}
In order to understand the effects of the interstitial element $X$ on 
magnetic properties, we have to understand the basic properties of 
the electronic structure associated with $X$, particularly $X$-$2p$ states. 
Figure~\ref{fig:atomic_level} shows 
the atomic $2s$ and $2p$ levels for each of $X$ 
from B to F together with the spin-unpolarized $3d$ and $4s$ levels of Fe. 
Although $2s$ levels are deeper than the Fe-$3d$ level for all $X$, 
$2p$ levels of B and C are definitely higher than the Fe-$3d$ level and 
the N-$2p$ level is nearly degenerate with the Fe-$3d$ level 
as pointed out by Kanamori.~\cite{Ka1990} 
In order to analyze the nature of the $X$-$2p$ states, 
most of the available theoretical papers take account of only the hybridization 
between $X$-$2p$ states and Fe-$3d$ states. 
However, it is important to take account of the Fe-$4s, \, p$ states 
which form broad $sp$ bands overlapping with the Fe-$3d$ bands 
in NdFe$_{11}$Ti$X$ like in ordinary transition metals.~\cite{Te1977} 
In the following discussion, 
we use the local coordinate as shown in Fig.~\ref{fig:relativeaxis_fe8j}. 
The $z$ axis is parallel to the crystal $c$ axis.

The B-$s$ and Fe(8$j$)-$s$, -$p$, -$d$ components of 
the partial DOS for NdFe$_{11}$TiB are shown in Fig.~\ref{fig:dos_splitoff}. 
(For other $X$ elements, see Figs.~S2 and S3 of
the Supplemental Material~\cite{supplementalmaterial}.)
We take a projection of the eigenstates within an atomic sphere 
with the radius $r_{c}$. 
(See Fig. S1 of the Supplemental Material~\cite{supplementalmaterial}.) 
A split-off band at around $-$8 eV is present in the B-$s$ DOS and 
a significant weight exists also near the bottom of the band above $-$7 eV. 
Such a feature is correlated with the feature existing 
in the $s$-component DOS of the Fe(8$j$)s which are the atoms closest to $X$. 
[Note that all partial DOSs associated with Fe(8$j$) are 
for four Fe(8$j$) atoms.
The local $x$ axis is in the direction from $X$ to each Fe(8$j$) atom.]
There is less significant weight of the $p$ and $d$ states of Fe(8$j$) 
in the energy range of the B-$2s$ band. 
Therefore, the states in the split-off B-$2s$ band and near the bottom of 
the continuous band are dominantly formed by the hybridization between 
B-$2s$ and Fe-$4s$ states. 
It is also important to note that there is no clear evidence of 
the antibonding states between B-$2s$ and Fe-$4s$ states 
pushed up above the Fermi level. 
This implies that one state is added as the occupied state 
for each spin state by introducing B to NdFe$_{11}$Ti. 
The split-off $X$-$2s$ band becomes deeper 
for higher valence elements, C to F, and 
the $2s$ band introduces one additional occupied state per spin for all $X$. 
This is an important result in the later discussion.

The character of the states in NdFe$_{11}$Ti$X$ associated with 
the $X$-$2p$ states needs more careful analysis. 
The overall features of the partial DOS for $X$-$2p$ states 
across the typical elements $X$ are shown in Figs.~S2 and S3 
in the Supplemental Material~\cite{supplementalmaterial}. 
For $X$-$2p_{x}$ and $2p_{y}$ states, 
both $\sigma$- and $\pi$-type hybridizations are possible with 
Fe(8$j$) states. 
We begin the discussion with the $\sigma$-type hybridization because it is stronger. 
For the pair of $X$ and Fe(8$j$)$_{0}$ located along the $x$ direction, 
the $2p_{x}$ state of $X$ and $s$, $p_{x}$, $d_{3z^{2}-r^{2}}$ and 
$d_{x^{2}-y^{2}}$ states of the Fe(8$j$)$_{0}$ are involved 
(Fig.~\ref{fig:dos_x-sigma}). 
The sharp structures seen in the $X$-$2p_{x}$ DOS are 
reflected in the DOSs of these states of Fe(8$j$) 
more or less depending on the strength of the hybridization. 
However, we see that the structure at the band bottom can also be seen 
in the DOSs of $p_{y}$ and $d_{xy}$ of Fe(8$j$) 
which do not form $\sigma$-type hybridization with $X$-$2p_{x}$. 
Such a band bottom structure cannot be understood in terms of 
$\pi$-type hybridization because it is even stronger 
for $d_{xy}$ than for $d_{x^{2}-y^{2}}$. 
The dominant mechanism of producing the structure is 
due to the following $\sigma$-type hybridization path; 
i.e., $2p_{y}$ at $X$ hybridizes strongly with $s$ of Fe(8$j$)$_{1}$ 
and Fe(8$j$)$_{3}$ of Fig.~\ref{fig:relativeaxis_fe8j} and then 
these $s$ states hybridize with $p_{y}$ and $d_{xy}$ states of Fe(8$j$)$_{0}$. 
On the other hand, 
the structures in the $X$-$2p_{z}$ component is of $\pi$-type origin 
and the $s$ states at Fe(8$j$) do not take part in the hybridization 
because four Fe(8$j$) atoms and the $X$ atom form a plane perpendicular to 
the crystal $c$ axis which is taken as the $z$-direction. 
Fe(8$j$)$_{0}$-$d_{zx}$ states and also the Fe(8$j$)$_{0}$-$p_{z}$ state form 
the $\pi$-type hybridization with the $X$-$2p_{z}$ state. 

From Figs.~\ref{fig:dos_x-sigma} and \ref{fig:dos_x-pi}, 
one may think that the $X$-$2p$ components near the bottom of 
the host band including the split-off bands may be characterized as 
the $X$-$2p$ dominating states. 
However, as pointed out above, for B and C at least, 
the $2p$ levels are located above the Fe-$3d$ levels. 
Therefore, the structures near the band bottom correspond to the bonding states 
among $X$-$2p$, Fe-$3d$, and (for $\sigma$) Fe-$4s,p$ states and 
the weight of B-$2p$ component near the band bottom is small 
as can be seen in Figs.~\ref{fig:dos_x-sigma} and \ref{fig:dos_x-pi}. 
In this energy range, both the Fe(8$j$)-$3d$ and $4s$ components for 
the $\sigma$ type and only Fe(8$j$)-$3d$ components for 
the $\pi$ type have larger weight. 
As we move from B to F, the partial DOS of the $X$-$2p$ component shifts to 
a lower energy and the weight increases. 
The difference between $\sigma$ type 
and $\pi$ type in the energy region of the bonding states 
comes from the difference in the hybridization strength. 
The $pd \sigma$ hybridization is stronger than the $pd \pi$ hybridization by 
$\sqrt{3}$ in the canonical band picture~\cite{An1975}
and moreover 
Fe(8$j$)-$4s$ states do not contribute to the $\pi$-type hybridization. 

The behavior of the antibonding states is quite different 
between the $\sigma$ bond and the $\pi$ bond. 
For the $\sigma$ bond, significant weight is 
present a bit higher than 6 eV from the Fermi level for B and 
this weight comes down to a lower energy for C and N. 
The structure becomes weak for O and nearly disappears for F. 
This is due to the deeper $2p$ levels and shrinking of 
the $2p$ wave functions for O and F. 
Because of the involvement of Fe(8$j$)-$4s$ states in the $\sigma$ bond, 
continuous spectra exist between the sharp structures of 
bonding and antibonding states. 
For B and C, however, 
the weight of the continuous spectra below the Fermi level is 
small and only the bonding states are occupied. 
Moving to N and O, 
we see that the weight of the continuous spectra particularly 
in the majority-spin band increases slightly for the $X$-$2p_x$, $2p_y$ states even below the Fermi level. 
Therefore, not only the bonding states but also some part of 
the antibonding states are occupied. 
The situation is different for the $\pi$ bond. 
As Fe(8$j$)-$4s$ states are not involved in this case, 
the antibonding states are just above the Fermi level already for B, 
about to be occupied for C, and occupied in the majority-spin band for N. 
Up to N, the $\pi$ antibonding state is above the Fe $d$ band and sharp. 
However, for O and F, the antibonding state moves into the $d$ band and 
becomes broad and small.
The antibonding states start to be occupied also in 
the minority-spin state for O and even more occupied for F. 
This behavior of the antibonding $\pi$ state can be clearly seen in 
$X$-$p_{z}$ and Fe(8$j$)-$d_{zx}$ in Fig.~\ref{fig:dos_x-pi}. 
Comparing these two DOSs, we find that 
the main component of the $\pi$ antibonding state is Fe(8$j$)-$d_{zx}$. 
Traces of the antibonding $\pi$ state can also be seen 
for Nd-$p_{z}$, $d_{3z^{2}-r^{2}}$ states as shown in Fig.~\ref{fig:dos_x-pi} .
The difference between C and N in the filling of the antibonding states 
in the majority-spin band was 
pointed out before for $R$Fe$_{12}X$~\cite{AsIsFu1993} and 
also for $R_2$Fe$_{17}X_{3}$.~\cite{StRiNiEs1996,AsYa1997}
We note that in both $R$Fe$_{12}X$ and $R_{2}$Fe$_{17}X_{3}$, 
$X$ and its nearest neighbor Fe atoms form a plane perpendicular to 
the direction from $X$ to $R$ and 
that the $s$ states at the Fe sites do not hybridize with 
the $X$-$2p_{z}$ state. 
Therefore, the filling of the $p_{\pi}$ antibonding states occurs for $X$=N 
in both systems. 

We make brief comments on the role of Nd in the $X$-$2p$ related states. 
We find some clear structures related to 
the bonding and antibonding $X$-$2p$ states but their weight is quite small. 
Based on this observation, we conclude that Nd does not play an important role 
in the bonding with the interstitial element $X$ and that 
the occupied antibonding state just below the Fermi level for N is ascribed to 
the hybridization between $X$-$2p_{z}$ and Fe(8$j$)-$3d_{zx}$ orbitals. 
Note, however, that the hybridization between 
$X$-$2p_{z}$ and Nd-$6p_{z}$, $5d_{3z^{2}-r^{2}}$ orbitals plays an important role 
in the crystal field parameter $A_{2}^{0}$ as described later. 

\subsection{Optimized Structures}
\begin{figure}[ht]
  \begin{center}
    \includegraphics[width=\hsize]{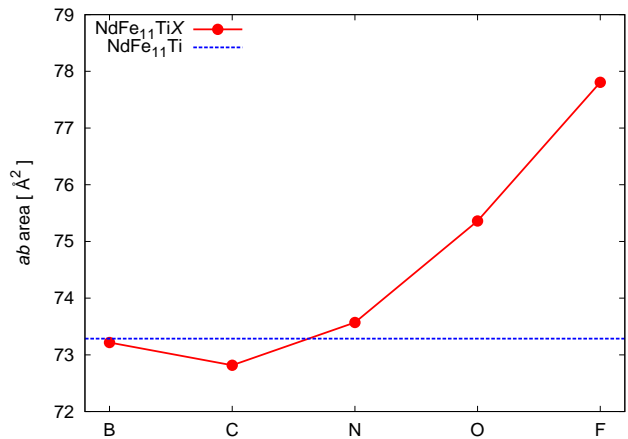}
  \end{center}
  \caption{(Color online) 
    $ab$ area of NdFe$_{11}$Ti$X$ ($X$=B, C, N, O, F) and NdFe$_{11}$Ti.}
  \label{fig:ab-area}
\end{figure}
As Table~\ref{table:latticeconstant} shows, 
the volume expands with the interstitial $X$, 
which leads to appreciable magnetovolume effects as shown later. 
We also observe the following interesting aspect 
in Table~\ref{table:latticeconstant} which reflects the bonding character 
related to the $X$-$2p$ states discussed above. 
Figure~\ref{fig:ab-area} shows the $X$ dependence of the $ab$ area. 
The broken horizontal line in this figure is the $ab$ area for NdFe$_{11}$Ti. 
For $X$=B, although the volume expands by about 5 \%, 
the $ab$ area decreases very slightly. 
The $ab$ area shrinks farther for $X$=C. 
Then as $X$ moves from C to F, the $ab$ area starts to increase. 
This trend is in clear contrast to the trend of the atomic radius 
which decreases monotonically 
from 0.88 {\AA} for B to 0.58 {\AA} for F.~\cite{Py2012}
The behavior of the $ab$ area of Fig.~\ref{fig:ab-area} reflects 
the strength of the covalent bond between 
$X$-$2p$ orbitals and Fe(8$j$)-$3d$, $4s$ orbitals. 
As was explained in the preceding section, 
only their bonding states are occupied for $X$=B and C. 
Therefore, the covalent bond between B or C and Fe(8$j$) tries to keep 
the B, C-Fe(8$j$) distance short. 
Moving from B to C, we notice that the bonding states become deeper and 
that the weight at C increases. 
These features contribute to a stronger covalent bond between C and Fe(8$j$) 
because the antibonding states are still mostly unoccupied. 
The situations for B and C qualitatively explain the variation of the $ab$ area 
in Fig.~\ref{fig:ab-area}. 
On the other hand, for $X$=N, the antibonding $2p_{z}$ state becomes occupied 
in the majority-spin band. 
At the same time, the tail of the $\sigma$-type antibonding states are 
also partly occupied. 
Therefore the covalent bond between N and Fe(8$j$) is weakened and 
the $ab$ area becomes larger. 
As we move to O and F, the antibonding states will be further occupied, 
$2p$ levels become much deeper, and $2p$ orbitals shrink in space. 
These features weaken further the $X$-Fe(8$j$) covalent bond and 
lead to further increase of the $ab$ area. 

\subsection{Magnetic Moment}
\begin{figure}[ht]
  \begin{center}
    \includegraphics[width=0.95\hsize]{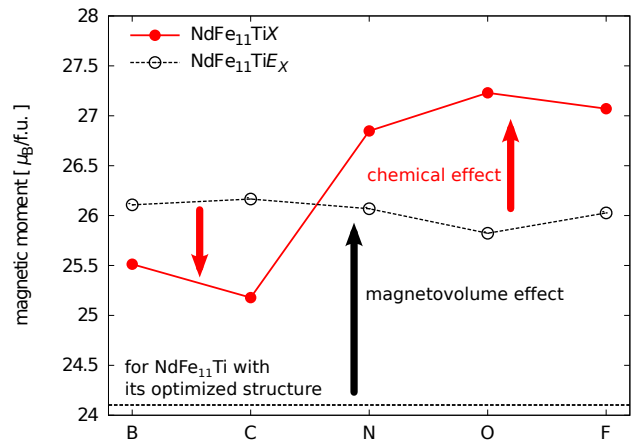}
  \end{center}
  \caption{(Color online) 
    The magnetic moment per formula unit for 
    NdFe$_{11}$Ti$X$ ($X$=B, C, N, O, F) and NdFe$_{11}$Ti.}
  \label{fig:mag_w-wo-x}
\end{figure}
\begin{figure}[ht]
  \begin{center}
    \includegraphics[width=\hsize]{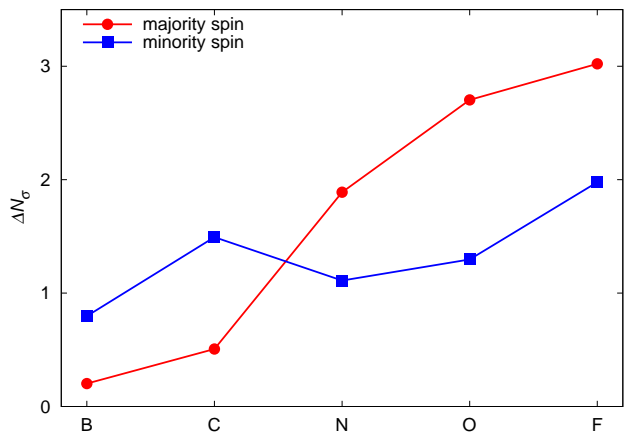}
  \end{center}
  \caption{(Color online) 
      The difference of the spin-resolved number of electrons between 
      NdFe$_{11}$Ti$X$ and NdFe$_{11}$Ti$E_{X}$. 
      Those of the majority-spin states and minority-spin states 
      are shown as the filled red circles and blue squares, respectively. 
  }
  \label{fig:majority-minority-spin-density}
\end{figure}
\begin{figure*}[ht]
  \begin{center}
    \includegraphics[width=0.8\hsize]{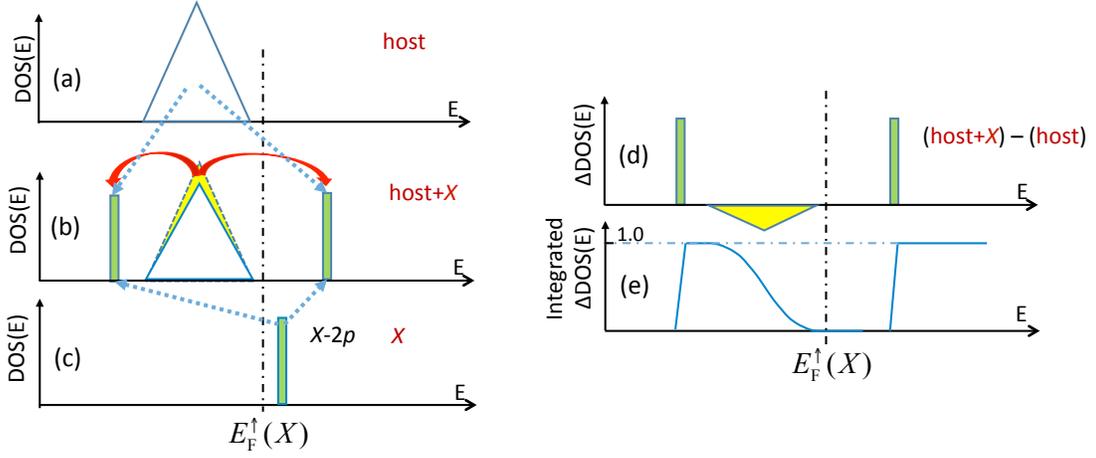}
  \end{center}
  \caption{(Color online) 
    A schematic picture for explaining the behavior of 
    $\Delta N_{\uparrow}(X)$ for $X$=B and C.}
  \label{fig:dos_schematic}
\end{figure*}
\begin{figure*}[ht]
  \begin{center}
    \includegraphics[width=\hsize]{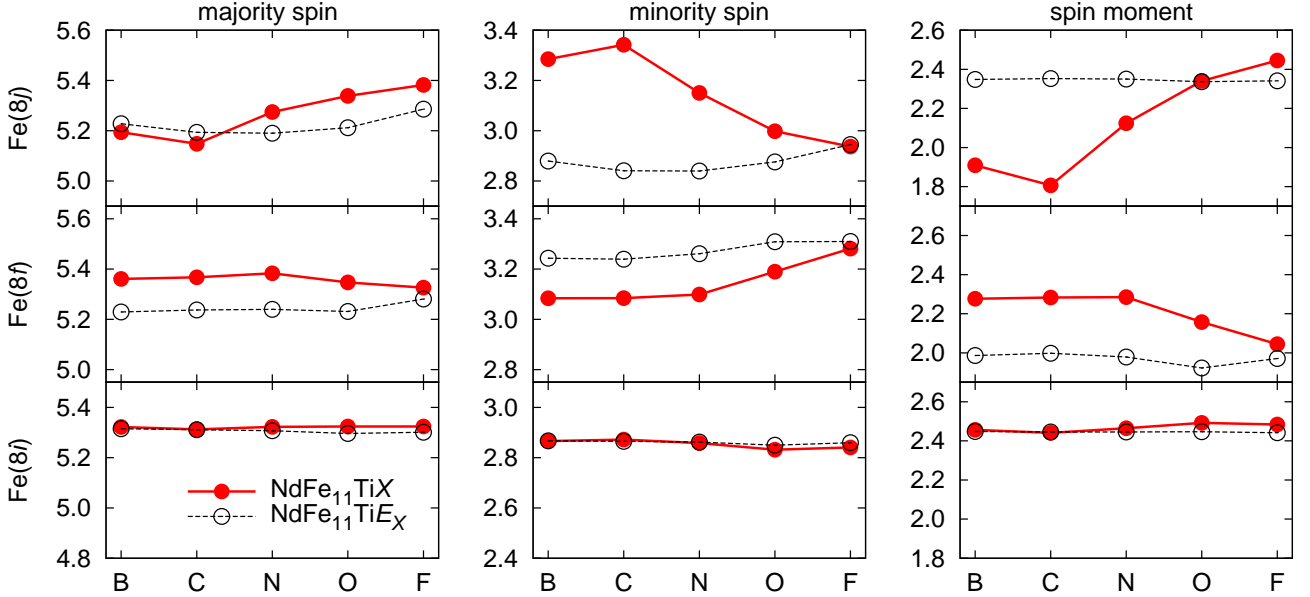}
  \end{center}
  \caption{(Color online) 
    The number of local majority-spin (left) and 
    minority-spin electrons (middle) and 
    the spin moment (right) at Fe sites 8$j$, 8$f$, and 8$i$ of 
    NdFe$_{11}$Ti$X$ and NdFe$_{11}$Ti$E_{X}$. 
    The integration region is given by a sphere with a radius 
    giving the same volume of each Bader region.}
  \label{fig:local-spin-fe}
\end{figure*}
We show by solid circles in Fig.~\ref{fig:mag_w-wo-x} 
the magnetic moment per formula unit for 
NdFe$_{11}$Ti$X$ for the optimized structures. 
The broken line near the bottom of the figure shows the magnetic moment of 
NdFe$_{11}$Ti with its optimized structure and the open circles near 
the middle of the figure show the magnetic moment of NdFe$_{11}$Ti$E_{X}$. 
The change in the magnetic moment from the bottom broken line to 
the open circles corresponds to the magnetovolume effect, 
which is associated with the expanded volume of NdFe$_{11}$Ti$E_{X}$ 
compared with that of NdFe$_{11}$Ti by 4 to 7 \% depending on $X$. 
However, among the open circles, the variation in the magnetic moment 
does not reflect the variation in the unit cell size. 
On the other hand, the change in the magnetic moment from the open circles to 
the solid circles corresponds to the chemical effect, 
which is the main subject in this section. 
Such a magnetovolume effect and a chemical effect in materials such as 
Fe$_{4}$N~\cite{AkTaTaKa1995}, 
$R$Fe$_{12-x}M_{x}X$~\cite{Sa1992,AsIsFu1993,AsYa1997,YaMaYaGeCh1997}, 
and $R_{2}$Fe$_{17}X_{3}$~\cite{UeHuFa1996,StRiNiEs1996,AsYa1997} 
systems were discussed before. 

In order to explain the trend in the chemical effect on 
the magnetic moment change, 
we show in Fig.~\ref{fig:majority-minority-spin-density} the change 
in the number of occupied states per unit cell 
for the spin state $\sigma$, $\Delta N_{\sigma}(X)$, caused by adding $X$ 
into NdFe$_{11}$Ti$E_{X}$. 
In this figure and also in the following argument, 
we use the following convention.
As already pointed out, the $2s$ states of $X$ form split-off bands 
below the host valence band and introduce two additional occupied states to 
the host band. 
Therefore, 
we assume the valence of $X$, $Z_{X}$, as 1, 2, 3, 4, and 5 
for B, C, N, O, and F, respectively, by focusing on only $2p$ states of $X$. 
Then $\Delta N_{\sigma}(X)$ should satisfy 
the charge neutrality condition given by
\begin{equation} \label{eq:charge}
 \Delta N_{\uparrow}(X)+ \Delta N_{\downarrow}(X)=Z_{X},
\end{equation}
On the other hand, the chemical effect part of the magnetic moment, 
$\Delta m$, in Fig.~\ref{fig:mag_w-wo-x} is given by
\begin{equation} \label{eq:delta_mag}
  \Delta m(X)=
  \Delta N_{\uparrow}(X)- \Delta N_{\downarrow}(X).
\end{equation}

There are a couple of important aspects 
in Fig.~\ref{fig:majority-minority-spin-density}. 
First, $\Delta N_{\uparrow}(X)$ is quite small for $X$=B and C. 
This can be understood in the following way.
To simplify the situation, we tentatively neglect 
the presence of wide continuous $4s$, $4p$ bands above 
the Fe-$3d$ bands and assume that the Fermi level 
in the majority-spin state is located in an energy region 
where there are no states. 
Such simplification is a good approximation as long as 
the $p_{\sigma}$ antibonding states 
do not make any significant contributions to the occupied states. 
Moreover, for the $p_{\pi}$ states, the contribution from the wide 
$4s$, $4p$ bands is quite small. 
Figure~\ref{fig:dos_schematic}(a) shows schematically the DOS of 
the host material. 
The Fermi level of the majority-spin state, 
$E_{\mathrm{F}}^{\uparrow}$, is located just above the host band. 
As shown in Fig.~\ref{fig:atomic_level}, B and C 
have their $2p$ levels located above the Fe-$3d$ band. 
With introducing the hybridization between the host states and 
the $X$-$2p$ state, 
a bonding state appears below the host band and the antibonding state above it
without being occupied. 
The bonding and antibonding states are formed 
from the original host states and the $X$-$2p$ state. 
The total number of the host states used to form 
the bonding and antibonding states is just one as long as the weight of 
the $X$-$2p$ state is one. 
Therefore, for the majority-spin state where the host band is fully occupied, 
while one additional state is introduced as the bonding state, 
the same number of states is subtracted from the host bands keeping 
the number of occupied states unchanged. 
This situation is shown in Figs.~\ref{fig:dos_schematic}(d) and \ref{fig:dos_schematic}(e). 
Figure~\ref{fig:dos_schematic}(d) shows the difference in the DOS 
between Fig.~\ref{fig:dos_schematic}(b) and Fig.~\ref{fig:dos_schematic}(a). 
The bonding and antibonding states are additionally introduced but 
some number of states whose total weight is one is subtracted 
from the host band. 
Figure~\ref{fig:dos_schematic}(e) shows the integrated DOS obtained 
by integrating the DOS of Fig.~\ref{fig:dos_schematic}(d) over energy. 
This quantity jumps up to unity when the upper limit of the integral crosses 
the bonding level, then starts to decrease as the upper limit of the integral 
moves into the host band region and reaches zero above the host band. 
It jumps up to unity again when the upper limit of the integral crosses 
the antibonding state. 
Therefore, as long as the antibonding state is unoccupied, 
the change in the number of occupied states per unit cell for 
the majority-spin state $\Delta N_{\uparrow}(X)$ is zero. 
At the same time, Fig.~\ref{fig:dos_schematic}(e) also shows that 
in the minority-spin band where 
the Fermi level is located within the host band, 
$\Delta N_{\downarrow}(X)>0$. 
With the above simplified model, 
we obtain that $\Delta N_{\uparrow}(X)=0$ for $X$ = B, C and 
$\Delta N_{\downarrow}(X)$=1 for $X$=B and 2 for $X$=C 
to satisfy Eq.~(4).
Accordingly, Eq.~(5) gives as the chemical effect part of the magnetic moment
$\Delta m(X)=-1$ for $X$=B and $-2$ for $X$=C. 
These results are qualitatively consistent with the behavior seen 
in Figs.~\ref{fig:mag_w-wo-x} and~\ref{fig:majority-minority-spin-density} 
for $X$=B and C. 
The deviation of the results in these figures 
from those of the simplified model comes from the partial filling of 
the antibonding states through the wide continuous $4s$ and $4p$ bands 
above the host band.

Second, the sudden increase in $\Delta N_{\uparrow}(X)$ 
from C to N is due to the filling of the antibonding $p_{\pi}$ state 
for the majority-spin state (Fig.~\ref{fig:dos_x-pi}) and 
leads to the sudden increase in $\Delta m(X)$ in Fig.~\ref{fig:mag_w-wo-x}. 
Further increase of $\Delta N_{\uparrow}(X)$ 
from $X$=N to F is due to the filling of $p_{\sigma}$ antibonding states. 
Third, compared with the behavior of $\Delta N_{\uparrow}(X)$, 
the variation of $\Delta N_{\downarrow}(X)$ is small. 
This is partly due to the fact that the Fermi level of 
the minority-spin state is located at the deep valley 
in the bulk density of states. 
To stabilize the band energy, the Fermi level is pinned in the valley region. 
Another important reason for the small variation of 
$\Delta N(E_{\mathrm{F}}^{\downarrow},X)$ is 
that the Fermi level of the minority-spin state is deep inside the $d$ band. 
This makes it difficult for the antibonding state to be occupied.
Therefore, the minority-spin $p_{\pi}$ antibonding state is filled
only for $X$=F though its weight is small due to the reduced $pd\pi$ hybridization mentioned before.

We now move to the discussion on the variation of the magnetic moment of 
Fe among 8$j$, 8$f$, and 8$i$ sites, which is shown in 
Fig.~\ref{fig:local-spin-fe}.
Because of the presence of Ti at one of the 8$i$ sites, 
the electronic structure is not the same among Fe atoms even within 
a given category of sites. 
Therefore, the average value is shown for each of 
8$j$, 8$f$, and 8$i$ sites. 
Kanamori explained the behavior seen in Fig.~\ref{fig:local-spin-fe} 
using the concept of $cobaltization$.~\cite{Ka1990}  
It says that by introducing B or C, 
the hybridization repulsion between $X$-$2p$ and Fe-$3d$ states pushes down 
the minority-spin $3d$ states of Fe(8$j$) more strongly than 
its majority-spin states leading to significant increase in 
the minority-spin electrons. 
Slight decrease in the majority-spin electrons is a result of 
the balance between the hybridization repulsion and 
the local charge neutrality requirement in metals, 
with the Fe(8$j$) being slightly more electron populated. 
The resulting electronic structure of Fe(8$j$) is similar to that of Co and 
the magnetic moment of Fe(8$j$) is reduced by about 0.4 $\mu_{\mathrm{B}}$. 
On the other hand, the change in the electronic structure at Fe(8$f$) is just 
opposite to that at Fe(8$j$). 
Kanamori also pointed out that the behavior at Fe(8$f$) can be explained 
in parallel with the change at the Fe sites next to Co in the Fe-Co alloy. 
The above picture of $cobaltization$ caused by the interstitial B is 
schematically explained by 
Figs.~1 and 2 in one of his papers.~\cite{OgAkKa2011} 
The concept of $cobaltization$ seems to be reasonable for $X$=B and C. 
However, the net magnetic moment is decreased by the chemical effect, 
and the enhancement by $X$ is due to the magnetovolume effect. 
For $X$=N, O, and F, the atomic $2p$ level is very close to or below 
the Fe-$3d$ level (Fig.~\ref{fig:atomic_level}) and 
the picture of $cobaltization$ is not applicable. 
The positive chemical effect for these elements is due mainly to 
coming down of the $p_{\pi}$ antibonding state 
below the Fermi level in the majority-spin state. 

\subsection{Crystal field parameter} \label{sec:crystalfieldparameter}
\begin{figure}[ht]
  \begin{center}
    \includegraphics[width=\hsize]{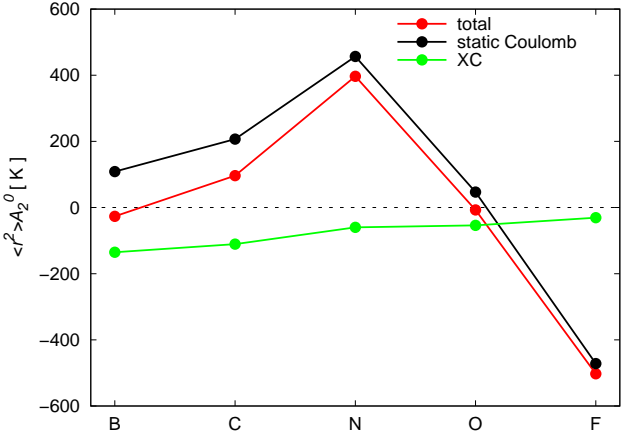}
  \end{center}
  \caption{(Color online) 
    $\langle r^{2} \rangle A_{2}^{0}$ of NdFe$_{11}$Ti$X$.
    Contributions from the static Coulomb and 
    the exchange-correlation potential are shown by 
    black circles and green circles, respectively. 
    Red circles show the sum of the two contributions.
  }
  \label{fig:cef_h-xc}
\end{figure}
\begin{figure}[ht]
  \begin{center}
    \includegraphics[width=\hsize]{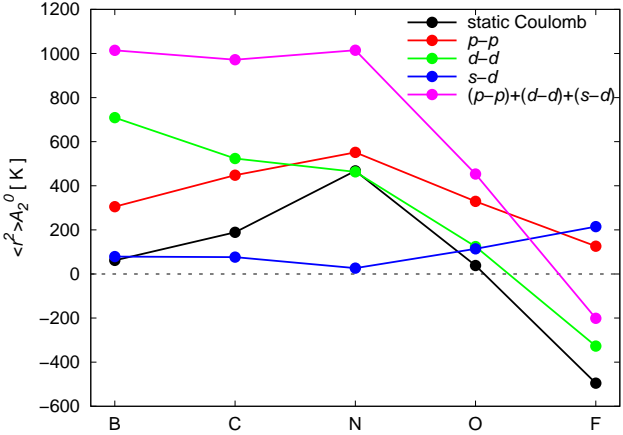}
  \end{center}
  \caption{(Color online) 
    $p$-$p$, $d$-$d$, and $s$-$d$ components in 
    $\langle r^{2} \rangle A_{2}^{0}$. 
    The cutoff radii are determined from Bader analysis. 
  }
  \label{fig:cef_s-p-d}
\end{figure}
\begin{figure}[ht]
  \begin{center}
    \includegraphics[width=\hsize]{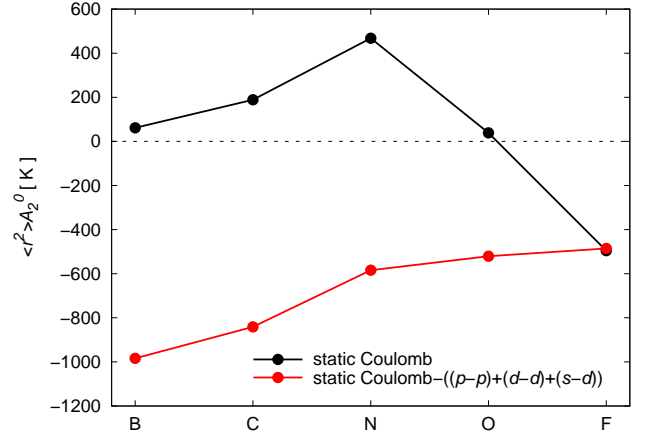}
  \end{center}
  \caption{(Color online) 
    Static Coulomb contribution to $\langle r^{2} \rangle A_{2}^{0}$ 
    brought by the off-site charge density (outside the Nd site). 
  }
  \label{fig:cef_ion}
\end{figure}
\begin{figure*}[th]
  \begin{minipage}{0.47\hsize}
    \includegraphics[width=\hsize]{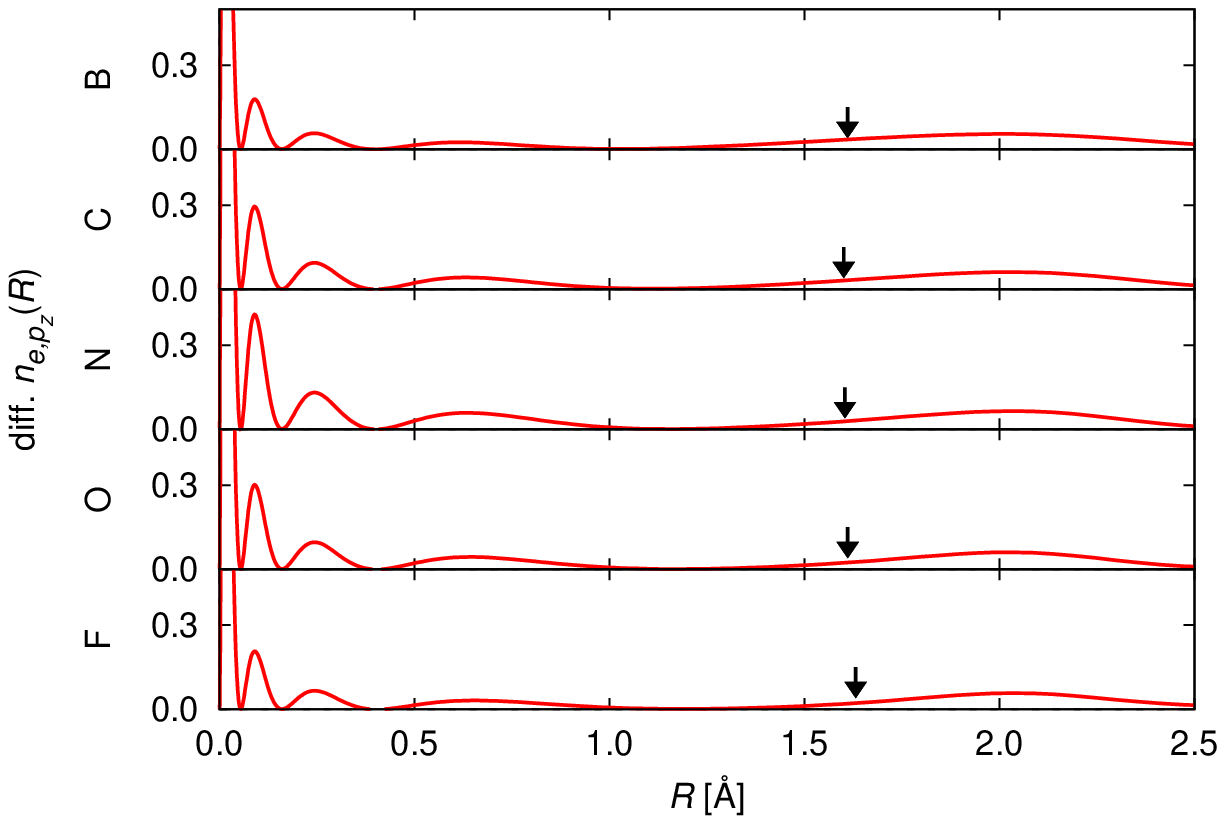}
  \end{minipage}
  \begin{minipage}{0.47\hsize}
    \includegraphics[width=\hsize]{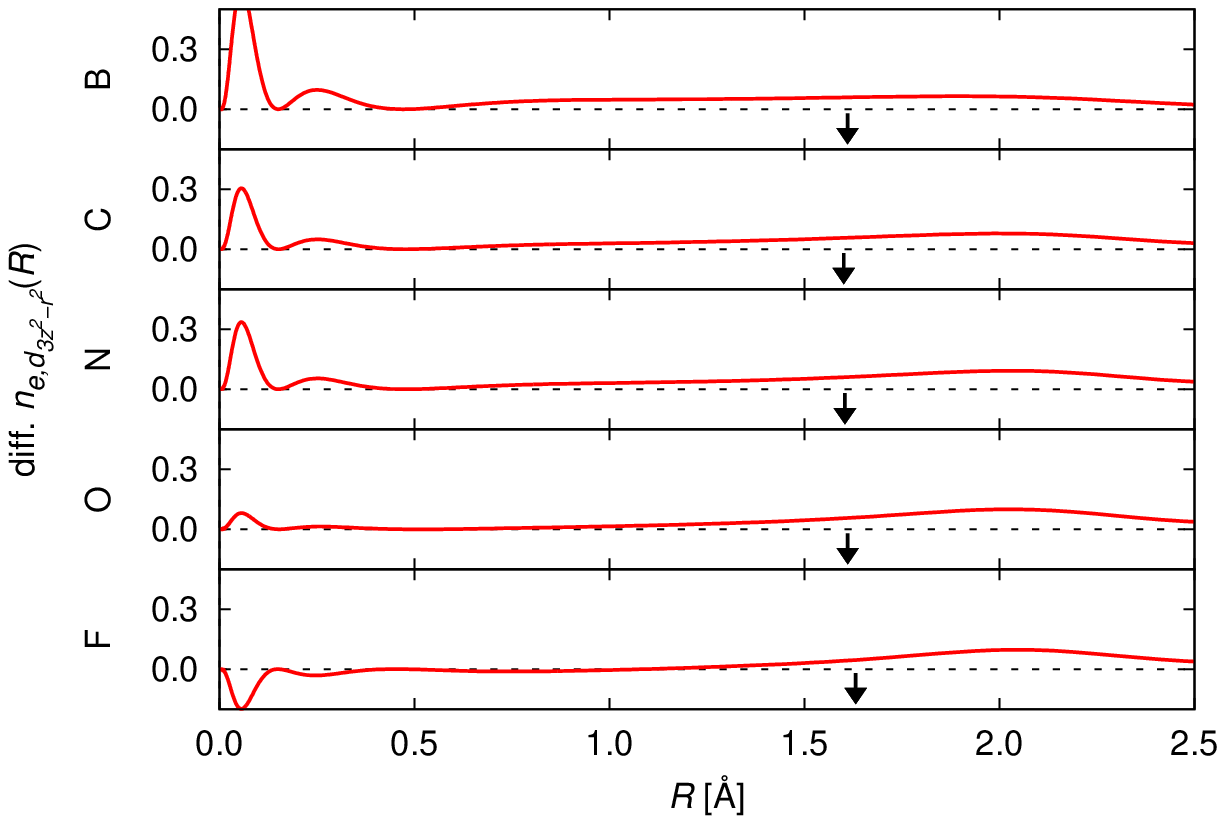}
  \end{minipage}
  \caption{(Color online) 
    $p_{z}$ and $d_{3z^{2}-r^{2}}$ components 
    ($n_{e,p_{z}} \equiv n_{e,1010}$ and $n_{e,d_{3z^{2}-r^{2}}} \equiv n_{e,2020}$)
    of the electron density difference 
    between NdFe$_{11}$Ti$X$ and NdFe$_{11}$Ti$E_{X}$ are shown 
    on the left and right panels, respectively. 
    The horizontal axis is the distance from Nd.
    The positive value means that 
    $n_{e,lml'm'}$ is increased by introducing $X$.
    The black arrows indicate the Bader radius.
  }
  \label{fig:rhoradial}
\end{figure*}
\begin{figure*}[th]
  \begin{minipage}{0.75\hsize}
    \begin{center}
      \includegraphics[width=\hsize]{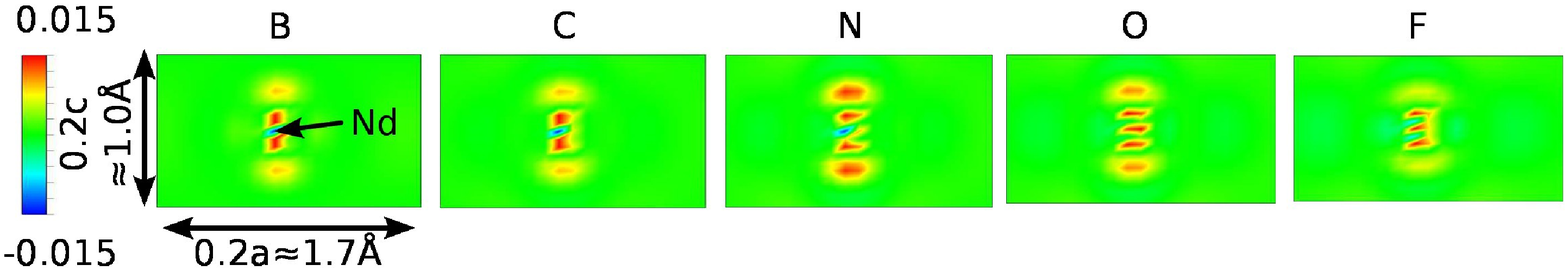}
    \end{center}
  \end{minipage}
  \begin{minipage}{0.20\hsize}
    \includegraphics[width=\hsize]{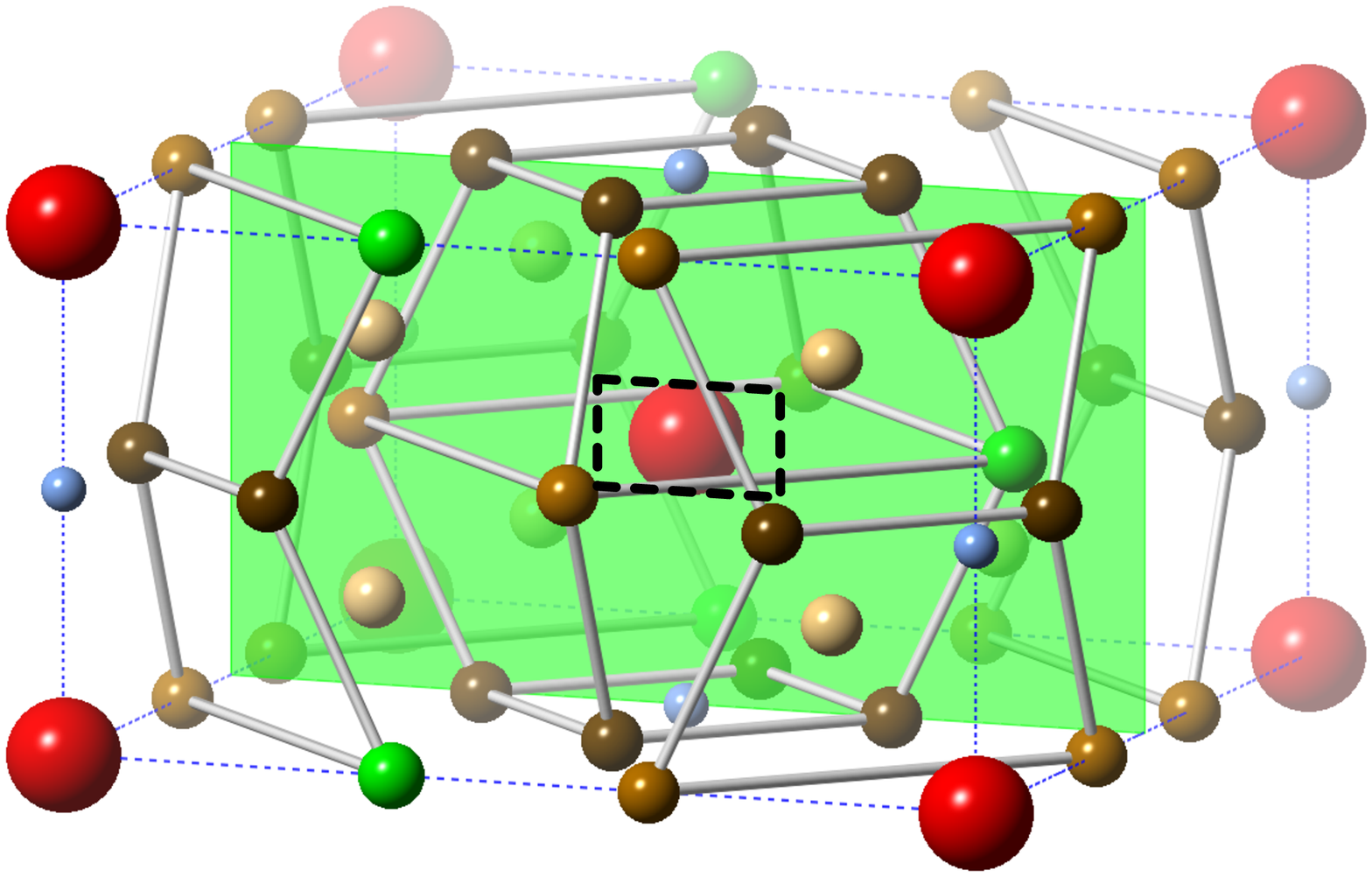}
  \end{minipage}
  \caption{(Color online) Electron density difference 
    between NdFe$_{11}$Ti$X$ and NdFe$_{11}$Ti$E_{X}$. 
    The density increases (decreases) in the red (blue) region. 
    The figures show the density difference map 
    on the $ac$ plane around the Nd atom. 
    The Nd atom is at the center of each figure. 
    The vertical and horizontal ranges are 0.2 times $c$ and $a$ axis, 
    respectively. 
    The displayed region is indicated by a black dashed box 
    on the right panel.
    The outermost red island corresponds to the peak at 0.25\AA\; 
    in Fig.~\ref{fig:rhoradial}.
}
  \label{fig:chargeplot}
\end{figure*}
Finally, we discuss the origin of 
the dopant dependence in $\langle r^{2} \rangle A_{2}^{0}$. 
In order to see 
whether the dopant dependence can be explained from the structure change, 
we plot $\langle r^{2} \rangle A_{2}^{0}$ of NdFe$_{11}$Ti$E_{X}$ 
in Fig.~\ref{fig:cef_total} by open circles. 
The difference among $X$=B, C, N, O, and F is within about 220 K and
the structural effect is not sufficient 
to explain the dopant dependence. 

The potential used in $\langle r^{2} \rangle A_{2}^{0}$ 
[Eq.~(\ref{eq:cef_veff})] 
consists of the static Coulomb (Hartree) and 
the exchange-correlation potentials. 
Figure~\ref{fig:cef_h-xc} shows 
the dopant dependence of each contribution. 
We can see that
the static Coulomb potential dominantly affects the dopant dependence of 
$\langle r^{2} \rangle A_{2}^{0}$. 
In the succeeding analysis, 
we concentrate on the contribution from the static Coulomb potential. 

The static Coulomb potential is caused by the charge density. 
We separate it into the on-site charge (at the Nd site) and 
off-site charge. 
Since the Nd-ion charge is spherically symmetric, 
the on-site contribution to $\langle r^{2} \rangle A_{2}^{0}$ is 
a consequence of the aspherical (e.g., $p$, $d$) component of 
the electron density at the Nd-site. 
To see this in more detail, 
we decompose the electron density within the Bader sphere at the Nd-site 
into the components having $Y_{lm}^{*}Y_{l'm'}$ as 
follows \cite{CoBuDiTh1990}.
{\setlength\arraycolsep{2pt}
  \begin{eqnarray}
    V_{\text{H}}(\Vec{r}) &=& -\int_{R<R_{c}} d^{3}R 
    \dfrac{\rho(\Vec{R})}{|\Vec{r}-\Vec{R}|}
    \label{eq:hartreepotential}
    \\
    \rho(\Vec{R}) &=& -\sum_{i}^{\text{occ}}
    \left|\sum_{l}\sum_{m=-l}^{m}f_{lm}^{i}(R)Y_{lm}(\hat{\Vec{R}})\right|^{2}
    \label{eq:electrondensity}
    \\
    &=& -\sum_{lml'm'} n_{e,lml'm'}(R)
    Y_{lm}(\hat{\Vec{R}})^{*}Y_{l'm'}(\hat{\Vec{R}})
    \label{eq:electrondensity_decomposed}
  \end{eqnarray}
}
where,
\begin{equation}
  n_{e,lml'm'}(R) \equiv \sum_{i}^{\text{occ}}
  f_{lm}^{i}(R)^{*} f_{l'm'}^{i}(R)
  \label{eq:electrondensity_decomposed_definition}
\end{equation}
$f_{lm}^{i}(R)$ is the radial part of 
the $lm$ component in the $i$th eigenstate. 
The cutoff radius $R_{c}$ of the integration in Eq.~(\ref{eq:hartreepotential}) 
is determined from the Bader region mentioned above (=$r_{c}$). 
[We note that $R_{c}$ is conceptually different from $r_{c}$, 
although the same value is taken in the actual calculations. The former is 
the cutoff radius of the charge density to evaluate the static Coulomb 
potential, whereas the latter is the cutoff radius for the potential 
in Eq.(\ref{eq:cef_veff}).] 
$1/|\Vec{r}-\Vec{R}|$ can be expanded by the real spherical harmonics:
\begin{equation} \label{eq:coulomb-sphericalharmonics}
  \frac{1}{|\Vec{r}-\Vec{R}|}=
  \sum_{lm}\frac{4\pi}{2l+1}\frac{r_{<}^{l}}{r_{>}^{l+1}}
  Z_{l}^{m}(\hat{\Vec{r}})Z_{l}^{m}(\hat{\Vec{R}})
\end{equation}
Here, $r_{<}\equiv\min(r,R)$ and $r_{>}\equiv\max(r,R)$. 
Inserting Eqs.(\ref{eq:electrondensity_decomposed}), 
(\ref{eq:electrondensity_decomposed_definition}), 
and (\ref{eq:coulomb-sphericalharmonics})
into Eq.(\ref{eq:hartreepotential}), 
and performing angular integration of 
$Z_{l}^{m}(\hat{\Vec{R}})Y_{l'}^{m'}(\hat{\Vec{R}})^{*}Y_{l''}^{m''}(\hat{\Vec{R}})$, 
three components, 
$\sum_{m=-1}^{1}n_{e,1m1m}$, $\sum_{m=-2}^{2}n_{e,2m2m}$, and $n_{e,0020}+n_{e,2000}$, 
yield nonzero contribution to $A_{2}^{0}$. 
The dopant dependencies of the three components, 
$p$-$p$ ($\equiv\sum_{m=-1}^{1}n_{e,1m1m}$), 
$d$-$d$ ($\equiv\sum_{m=-2}^{2}n_{e,2m2m}$), 
and $s$-$d$ ($\equiv n_{e,0020}+n_{e,2000}$), 
are shown in Fig.~\ref{fig:cef_s-p-d}.
Among the three, 
the $p$-$p$ electron density gives a peak structure at $X$=N, 
which is similar to the behavior of the total $\langle r^{2} \rangle A_{2}^{0}$. 
In contrast, 
the contribution from $d$-$d$ density decreases monotonically from $X=$B to F. 
The $s$-$d$ cross-term contribution is small 
compared with the $p$-$p$ and $d$-$d$ contributions. 
We also plot the remaining component defined as 
the static Coulomb minus 
the contribution from the electron density within the Bader atomic sphere 
in Fig.~\ref{fig:cef_ion}.
The contribution increases monotonically as the dopant changes from B to F.

In order to understand the dopant dependence of 
the $p$-$p$ and $d$-$d$ components, 
we decompose them into each $m$ component, and 
compare the results for NdFe$_{11}$Ti$X$ and for NdFe$_{11}$Ti$E_{X}$.
We found that 
most of the components are not changed 
between NdFe$_{11}$Ti$X$ and NdFe$_{11}$Ti$E_{X}$.
However, the $p_{z}$ and $d_{3z^{2}-r^{2}}$ components are significantly changed 
as shown in Fig.~\ref{fig:rhoradial}, where 
the $n_{e,lml'm'}(R)$ differences for $p_{z}$ and $d_{3z^{2}-r^{2}}$ 
are plotted as a function of the distance $R$ from Nd.
The changes are prominent in the vicinity of the Nd atom 
within a 0.5 \AA\; radius. 
In Fig.~\ref{fig:chargeplot}, the density difference map 
on the $ac$ plane including the Nd atom is shown.
Both $p_{z}$ and $d_{3d^{2}-r^{2}}$ orbitals have an anisotropic shape 
expanding along the $c$ axis.
Correspondingly, an increase in the density difference is observed 
above and below the Nd site along the $c$ axis. 
This accumulated electron charge suppresses 
the aspherical $4f$-electron density from above and below the Nd site, 
which induces the uniaxial anisotropy.
For $p_{z}$ density, 
the difference increases from $X$=B to N and decreases from N to F, and 
for $d_{3z^{2}-r^{2}}$, it decreases monotonically from $X$=B to F.
These density differences near the Nd site explain 
the dopant dependence of the $p$- and $d$-components. 
The influence of the bonding charge between $X$ and $R$ has also been discussed 
in Ref.~\onlinecite{AsYa1997}.

There still remains the contribution in the static Coulomb potential 
other than the contribution from the electron density 
within the Bader atomic sphere. 
This can be interpreted as the off-site contribution, namely 
the contribution from the charges outside the Nd-atomic sphere with $R_{c}$.
The increasing behavior of this contribution shown in Fig.~\ref{fig:cef_ion} 
reflects the electronegativity of the interstitial dopants.
The dopant elements B, C, N, O, and F have the electronegativity 
2.04, 2.55, 3.04, 3.44, 3.98, respectively. \cite{Al1961}
F tends to attract the negative charge more than B.
The attracted negative charge at the $X$ site gives 
the Coulomb repulsive interaction at the Nd site along the $c$ axis. 
This contribution for $\langle r^{2} \rangle A_{2}^{0}$ has a positive value. 
Thus, F gives a less negative value than B.

These three contributions, $p$-$p$, $d$-$d$ electron charges at Nd-site, 
and the charge at the interstitial dopant, explain 
the nonmonotonic dopant dependence of $\langle r^{2} \rangle A_{2}^{0}$. 

\section{Conclusion}
We have studied the magnetic properties of NdFe$_{11}$Ti$X$ with 
$X$=B, C, N, O, and F with first-principles electronic structure calculations to find a better interstitial dopant 
for a permanent magnet.
According to our study, the total magnetic moment is increased 
by doping B, C, N, O, and F at the interstitial site. 
The magnetic moment increase can be separated into the magnetovolume effect and the 
chemical effect, the latter of which depends
strongly on the dopant.  The chemical effect is negative for B and C and becomes suddenly positive for N, O, and F.
By studying carefully the basic electronic structures of NdFe$_{11}$Ti$X$ 
from the viewpoint of the hybridization between $X$ and Fe(8$j$), 
we gave an explanation to the mechanism of the negative chemical effect for B and C. 
The sudden increase in the chemical effect part of the magnetic moment 
from C to N is due to the coming down below the Fermi level of 
the majority-spin $p_{\pi}$ antibonding state formed between $X$ and Fe(8$j$).

The results for $\langle r^{2} \rangle A_{2}^{0}$ suggest that 
the strongest enhancement of uniaxial anisotropy is achieved 
by $X$=N. 
We analyzed the dopant dependence of $\langle r^{2} \rangle A_{2}^{0}$ from 
the decomposed electron density $n_{e,lml'm'}(R)$.
The analysis suggests that 
the bonding charge between Nd and $X$, and the charge at the $X$ site 
can explain the increasing trend from B to N and 
decreasing trend from N to F of the magnetocrystalline anisotropy. 

The present study suggests that
the interstitial nitrogenation is the most appropriate 
among the typical elements B, C, N, O, and F doping 
in terms of the magnetization and magnetocrystalline anisotropy.

\begin{acknowledgments}
The authors would like to thank 
Prof. H. Akai and 
Dr. S. Hirosawa 
for fruitful discussions.
This work was supported by 
the Elements Strategy Initiative Project under the auspices of MEXT, 
MEXT HPCI Strategic Programs for Innovative Research (SPIRE) and 
Computational Materials Science Initiative (CMSI). 
The computations have been partly carried out using the facilities of 
the Supercomputer Center, the Institute for Solid State Physics, 
the University of Tokyo, and the supercomputer of ACCMS, Kyoto University, and 
also by the K computer provided by the RIKEN Advanced Institute for 
Computational Science (Project ID:hp140150). 
\end{acknowledgments}

\bibliography{./Reference,comment}

\clearpage

\renewcommand{\thefigure}{S\arabic{figure}}
\setcounter{figure}{0}

\section{Supplemental Material}
In this Supplemental Material, we show the following figures.
\begin{enumerate}
\item Fig.~\ref{fig:radius_pdos}: The radius of the atomic sphere whose volume is that of the Bader region~\cite{Ba1990,HeArJo2006} for each atom in the unit cell of NdFe$_{11}$Ti$X$.  The same atomic sphere radius is used for NdFe$_{11}$Ti$E_X$. 
\item Fig.~\ref{fig:dos_x}: The partial DOS of $X$ in NdFe$_{11}$Ti$X$.
\item Fig.~\ref{fig:dos_x_zoom}: The zoomed figures of Fig.~\ref{fig:dos_x}.
\item Fig.~\ref{fig:dos_fe8j}: The partial DOS of Fe(8$j$) in NdFe$_{11}$Ti$X$.
\item Fig.~\ref{fig:dos_nd}: The partial DOS of Nd in NdFe$_{11}$Ti$X$.
\item Fig.~\ref{fig:spin_x}: The number of electrons for each spin state in the atomic sphere for each $X$ in NdFe$_{11}$Ti$X$.
\end{enumerate}

The partial DOSs in Figs.~\ref{fig:dos_x}, \ref{fig:dos_x_zoom}, 
\ref{fig:dos_fe8j} and \ref{fig:dos_nd} and 
the number of electrons in Fig.~\ref{fig:spin_x} are 
given for the atomic sphere with the radius given in Fig.~\ref{fig:radius_pdos}.
$X$-$s$ and $p_{x}+p_{y}$ components 
in Figs.~\ref{fig:dos_x} and \ref{fig:dos_x} are related with 
the $\sigma$ bond between $X$ and Fe(8$j$), 
while X-$p_z$ component is related with the $\pi$ bond.  
In Fig.~\ref{fig:spin_x}, the $X$-$p_{z}$ component shows 
a jump from $X$=C to N. 
This is due to the occupation of the anti-bonding states at $X$=N.

\clearpage

\begin{figure}[h]
  \begin{center}
    \includegraphics[width=\hsize]{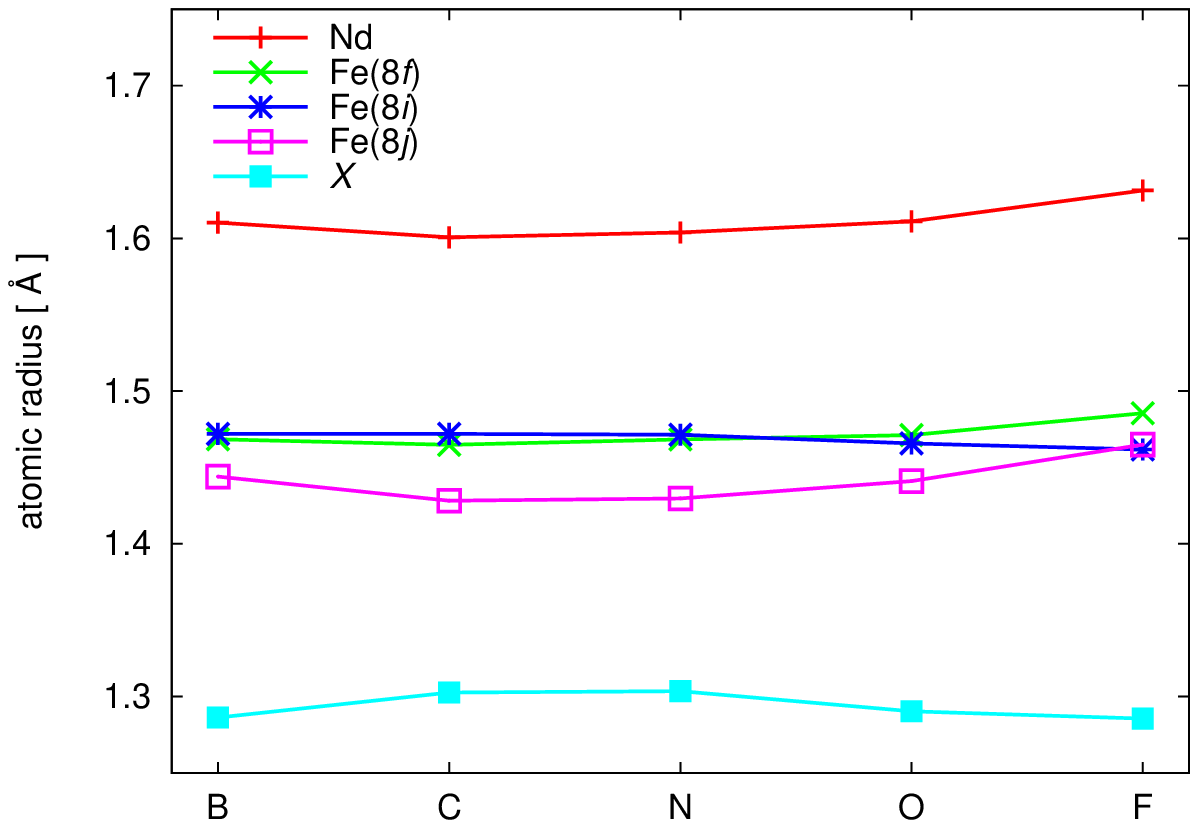}
  \end{center}
  \caption{(Color online)
    The sphere radius used for the integration region of the partial DOS for each $X$.
  }
  \label{fig:radius_pdos}
\end{figure}
\begin{figure*}[ht]
  \begin{center}
    \includegraphics[width=0.9\hsize]{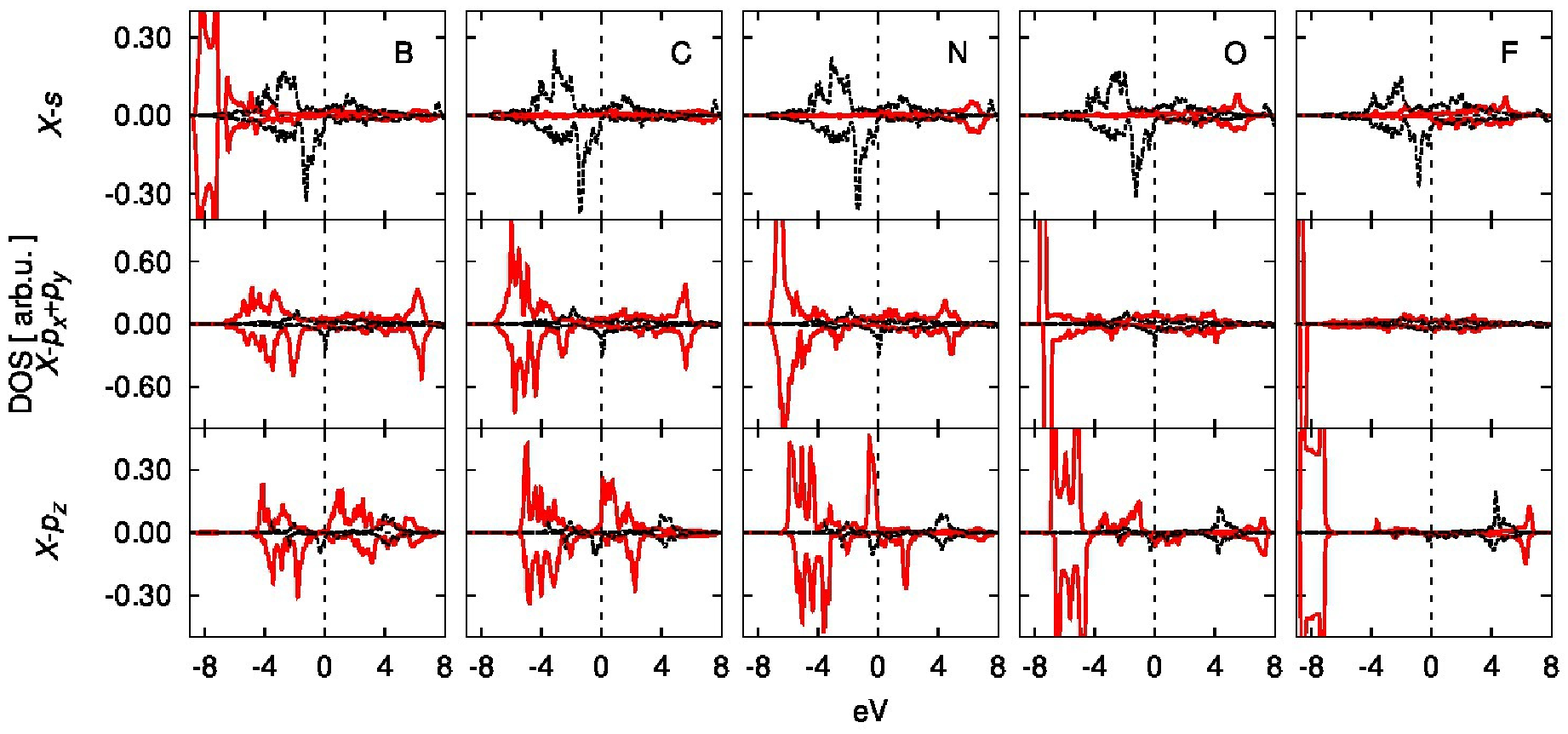}
  \end{center}
  \caption{(Color online) Partial density of states 
    at $X$ of NdFe$_{11}$Ti$X$.
    The $s$ components (upper panel), 
    the sumation of $p_{x}$ and $p_{y}$ components (middle panel), 
    and $p_{z}$ components (lower panel) are shown.}
  \label{fig:dos_x}
\end{figure*}
\begin{figure*}[ht]
  \begin{center}
    \includegraphics[width=0.9\hsize]{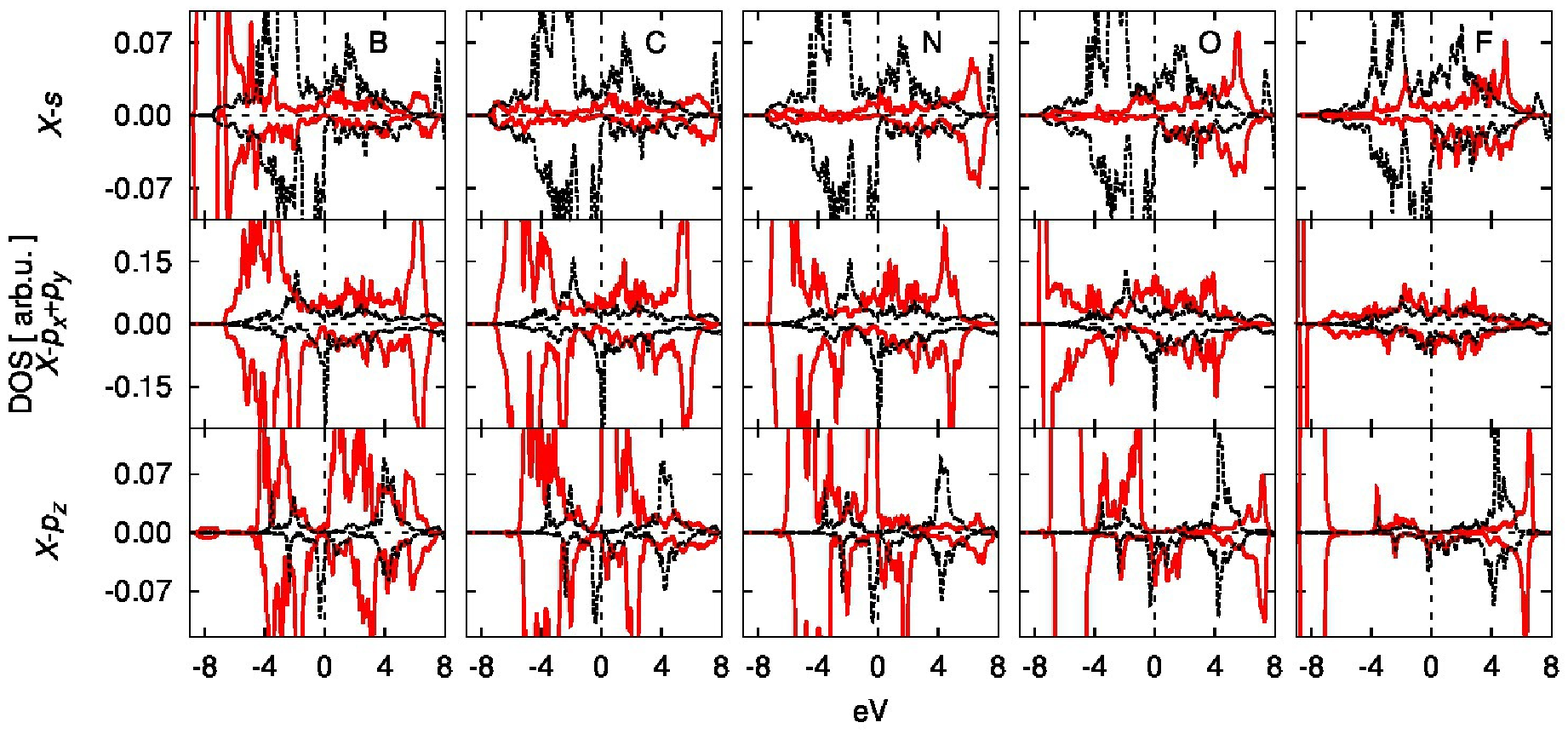}
  \end{center}
  \caption{(Color online) Zoomed figures for 
    partial density of states at $X$ of NdFe$_{11}$Ti$X$ 
    corresponding to Fig.~\ref{fig:dos_x}.
  }
  \label{fig:dos_x_zoom}
\end{figure*}
\begin{figure*}[ht]
  \begin{center}
    \includegraphics[width=0.9\hsize]{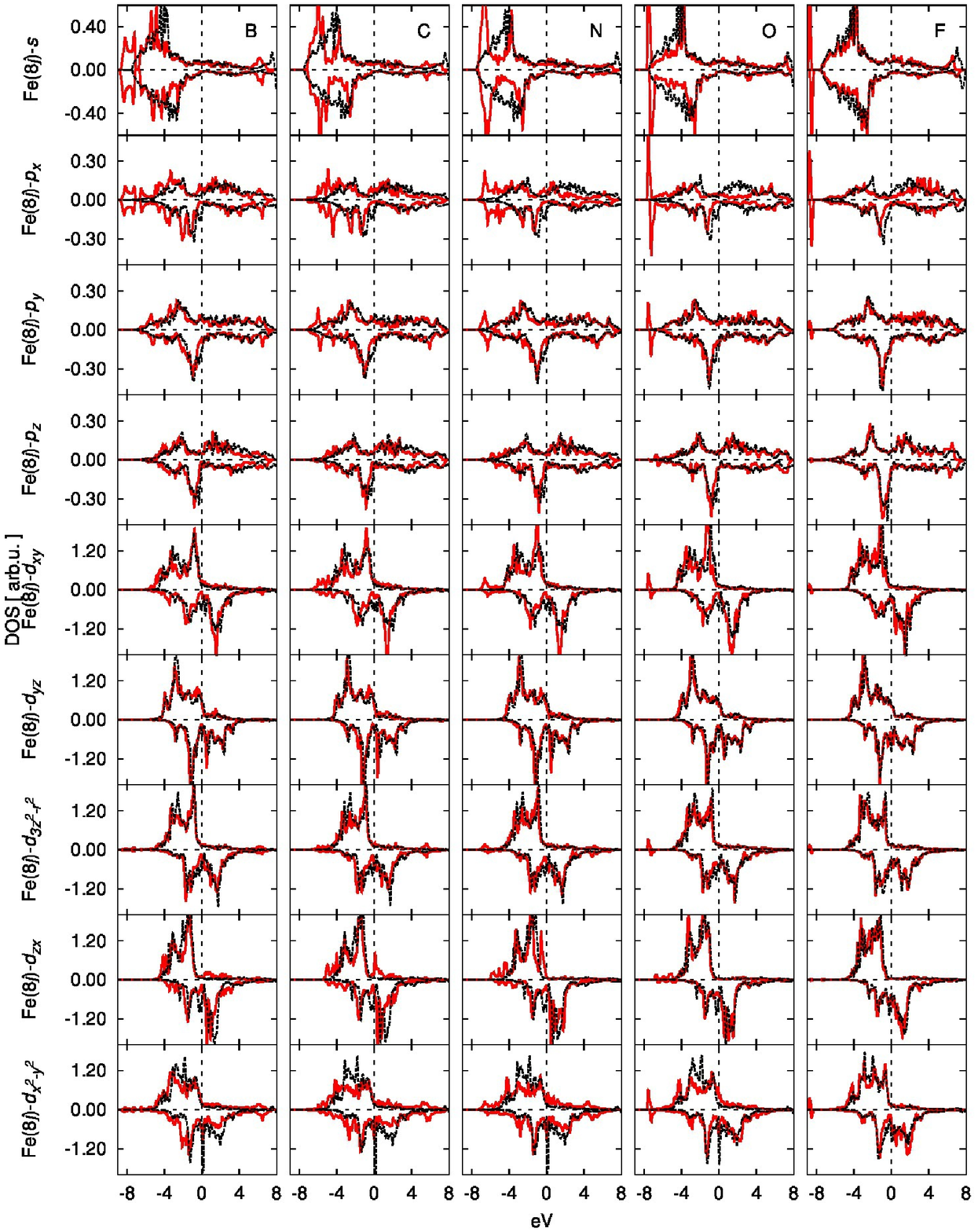}
  \end{center}
  \caption{(Color online) Partial density of states 
    at Fe(8$j$) of NdFe$_{11}$Ti$X$.
    The $s$, $p_{x}$, $p_{y}$, $p_{z}$, 
    $d_{xy}$, $d_{yz}$, $d_{3z^{2}-r^{2}}$, $d_{zx}$, $d_{x^{2}-y^{2}}$ components 
    are shown.  The red solid lines for NdFe$_{11}$Ti$X$ and the black broken lines for NdFe$_{11}$Ti$E_X$.
  }
  \label{fig:dos_fe8j}
\end{figure*}
\begin{figure*}[ht]
  \begin{center}
    \includegraphics[width=0.9\hsize]{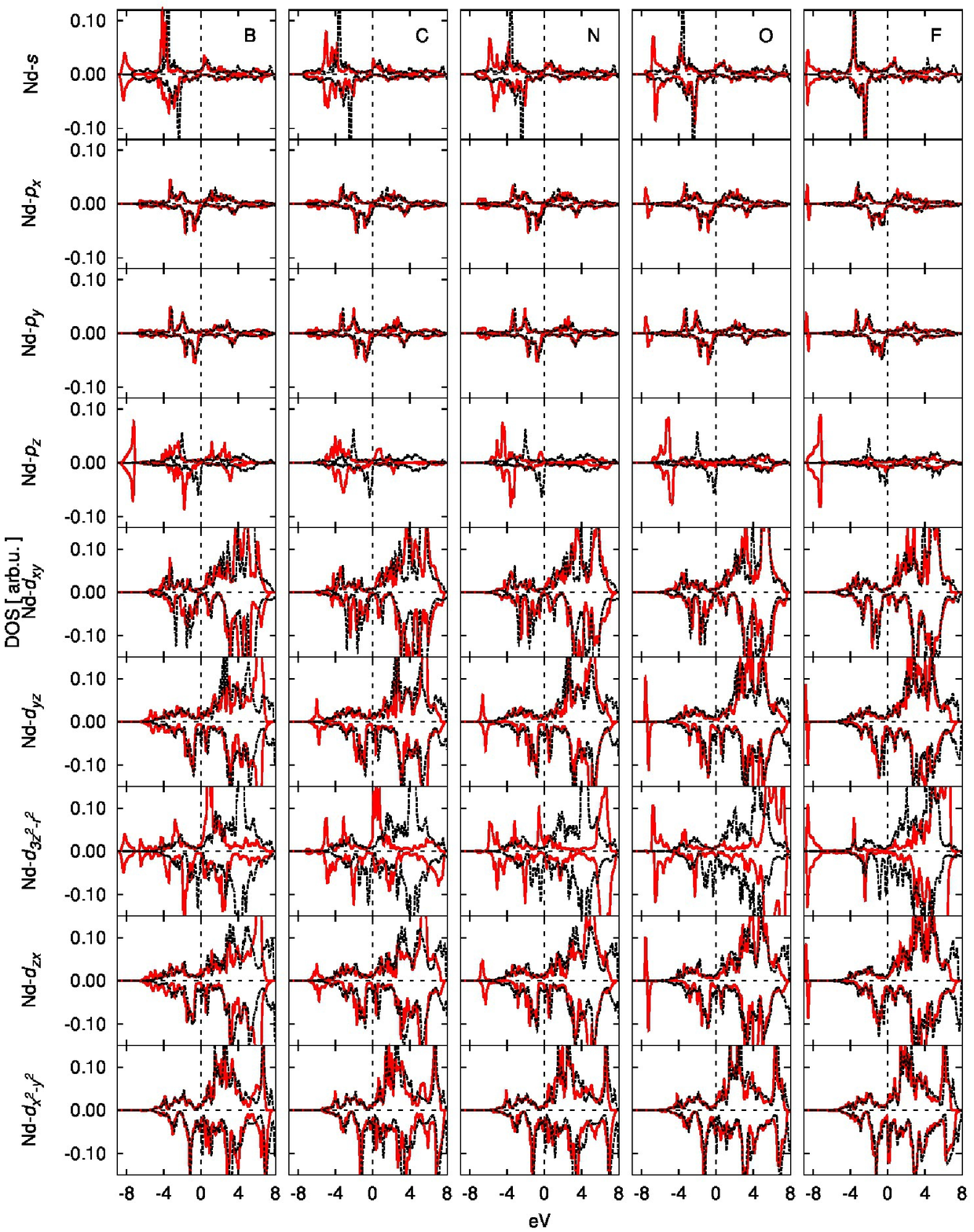}
  \end{center}
  \caption{(Color online) Partial density of states 
    at Nd of NdFe$_{11}$Ti$X$. 
    The $s$, $p_{x}$, $p_{y}$, $p_{z}$, 
    $d_{xy}$, $d_{yz}$, $d_{3z^{2}-r^{2}}$, $d_{zx}$, $d_{x^{2}-y^{2}}$ components 
    are shown.   The red solid lines for NdFe$_{11}$Ti$X$ and the black broken lines for NdFe$_{11}$Ti$E_X$.
  }
  \label{fig:dos_nd}
\end{figure*}
\begin{figure}[h]
  \begin{center}
    \includegraphics[width=\hsize]{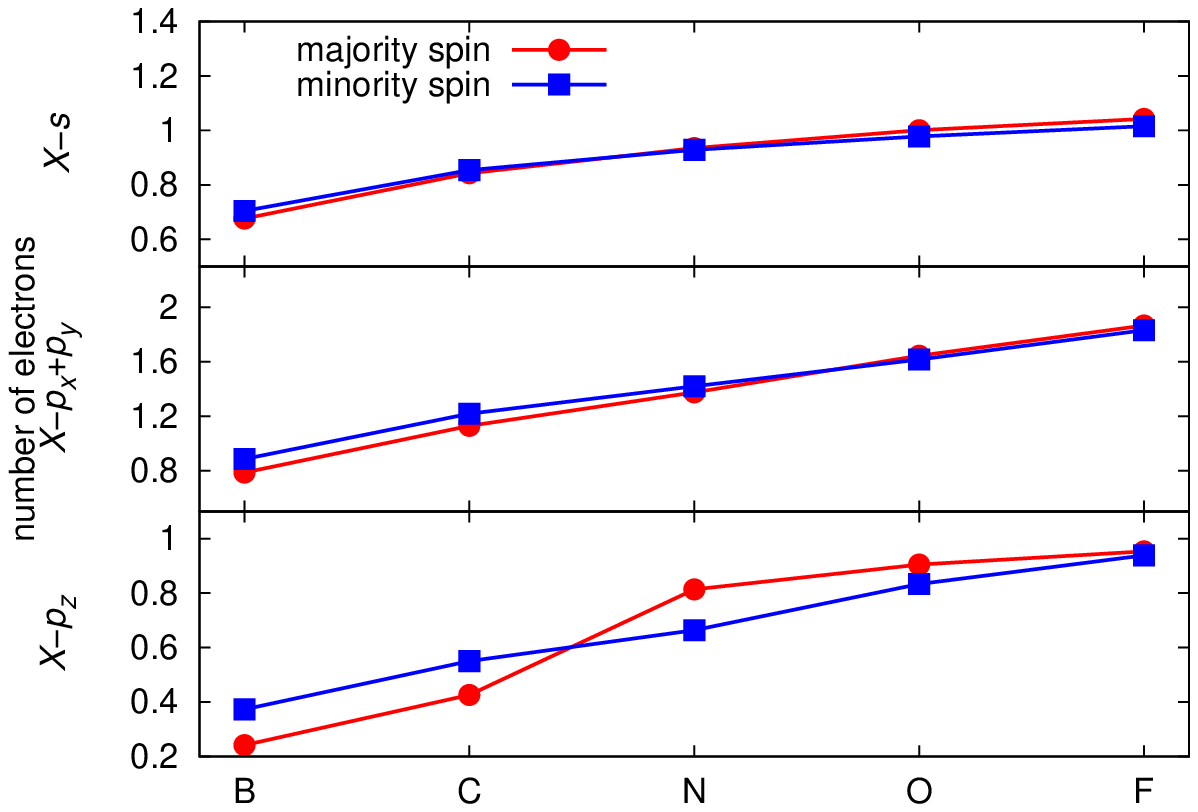}
  \end{center}
  \caption{(Color online)
    The number of majority spin and that of minority spin electrons 
    within the sphere around $X$. 
    The components are seperated into each orbitals, 
    $s$, $p_{x}+p_{y}$, and $p_{z}$. 
  }
  \label{fig:spin_x}
\end{figure}

\end{document}